\documentclass[10pt,twocolumn,preprintnumbers,amsmath,amssymb,nofootinbib,superscriptaddress,prd]{revtex4-2}

\usepackage[dvipsnames]{xcolor}
\usepackage{bm}
\usepackage{graphicx}
\usepackage[colorlinks,allcolors=RoyalBlue]{hyperref}
\usepackage[normalem]{ulem}
\newcommand{\eq}{\mathrm{eq}}

\newcommand{\fax}{f_{\rm ax}} 

\newcommand{\lfs}{\lambda_{\rm fs}}
\newcommand{\phip}{\phi_p}

\newcommand{\qaem}{\hat q}
\newcommand{\bk}{{\bf k}}
\newcommand{\bq}{{\bf q}}
\newcommand{\bx}{{\bf x}}
\newcommand{\br}{{\bf r}}
\newcommand{\bp}{{\bf p}}
\newcommand{\Schro}{Schr\"{o}dinger}
\newcommand{\Npix}{256}

\begin{document}
\preprint{FERMILAB-PUB-24-0296-T}
\title{Warm and Fuzzy Dark Matter: Free Streaming of Wave Dark Matter}

\author{Rayne Liu}
\affiliation{Kavli Institute for Cosmological Physics, Enrico Fermi Institute, and Department of Astronomy \& Astrophysics, University of Chicago, Chicago IL 60637
}

\author{Wayne Hu}
\affiliation{Kavli Institute for Cosmological Physics, Enrico Fermi Institute, and Department of Astronomy \& Astrophysics, University of Chicago, Chicago IL 60637
}

\author{Huangyu Xiao}
\affiliation{Kavli Institute for Cosmological Physics, Enrico Fermi Institute, and Department of Astronomy \& Astrophysics, University of Chicago, Chicago IL 60637
}
\affiliation{Theory Division, Fermi National Accelerator Laboratory, Batavia, IL 60510, USA}
\begin{abstract}
Wave or fuzzy dark matter that is produced with relativistic wavenumbers exhibits free streaming effects analogous to warm or hot particle dark matter with relativistic momenta.  Axions produced after inflation provide such a warm or mildly relativistic candidate, where the enhanced suppression and observational bounds are only moderately stronger than that from wave propagation of initially cold axions. More generally, the free streaming damping also impacts isocurvature fluctuations from generation in causally disconnected patches.  As coherent spatial fluctuations free stream away they leave incoherent and transient superpositions in their wakes.   These multiple wave momentum streams are the wave analogue of particle phase space fluctuations or directional collisionless damping of massive neutrinos or hot dark matter. The observable impact on both adiabatic and isocurvature fluctuations of fuzzy dark matter can differ from their cold dark matter counterparts due to free streaming depending on how warm or hot is their momentum distribution.
\end{abstract}

\date{\today}

\maketitle

\section{Introduction} 

Wave dark matter refers to bosonic dark matter with masses $m\lesssim 30$ eV such that the occupation number is much greater than one, and can arise in a variety of theoretical contexts~(see e.g.~\cite{Hui:2021tkt} for a review).
One of the leading candidates is axion dark matter, where dark matter behaves as a classical wave below the de Broglie scale.
The axion, originally proposed to explain upper limits on the neutron electric dipole moment and solve the strong CP problem \cite{Peccei:1977hh, PhysRevLett.40.223, PhysRevLett.40.279, PhysRevLett.43.103, Peccei:2006as}, is also a viable dark matter candidate \cite{Abbott:1982af, Dine:1982ah,Preskill:1982cy} and has stimulated much interest in dark matter physics and numerical experiment searches \cite{Sikivie:2013laa,Kahn:2016aff,Ouellet:2018beu,Gramolin:2020ict,Salemi:2021gck,Zhang:2021bpa,DMRadio:2022jfv,Jaeckel:2007ch,ALPS:2009des,Caspers:2009cj,Ehret:2010mh,Redondo:2010dp,Bahre:2013ywa,Betz:2013dza,OSQAR:2015qdv,Janish:2019dpr,Berlin:2019ahk,Berlin:2020vrk,Gao:2020anb,Berlin:2022hfx,ADMX:2018gho,ADMX:2019uok,Alesini:2019ajt,Lee:2020cfj,Alesini:2022lnp,DeRocco:2018jwe,Obata:2018vvr,Liu:2018icu,Fedderke:2023dwj}. The mass spectrum of axions extends beyond the original QCD axion for the strong CP problem \cite{Arvanitaki:2009fg}, and we will use the term  ``axion" for any light (pseudo)scalar dark matter that has similar interactions to the QCD axion. Another interesting candidate is the dark photon dark matter that can be produced from a variety of mechanisms \cite{Graham:2015rva,Agrawal:2018vin,Co:2018lka,Dror:2018pdh,Bastero-Gil:2018uel,Long:2019lwl,Co:2021rhi}. While we focus on ultralight axions in this work, similar physical phenomena often apply to the dark photon and axion dark matter in general.

 For wave dark matter on the higher mass end, its de Broglie wavelength is much shorter than astrophysical scales, and laboratory experiments are necessary. Once the wave dark matter is ultralight (often called fuzzy dark matter \cite{Hu:2000ke}), cosmological measurements on the linear power spectrum or stellar kinematics of ultrafaint dwarfs can constrain lower mass ranges \cite{Hu:2000ke,Hui:2016ltb,Irsic:2017yje,Kobayashi:2017jcf,Nori:2018pka,Schutz:2020jox,Dalal:2022rmp}. The wave nature of fuzzy dark matter can lead to rich phenomenology such as the formation of soliton cores at halo centers and interference effects \cite{Friedberg:1986tp,Friedberg:1986tq,Seidel:1993zk,Guzman:2006yc,Schive:2014hza,Veltmaat:2018dfz,May:2022gus}. Therefore, fuzzy dark matter can also be probed by compact objects through its wave dynamics \cite{Baryakhtar:2020gao,Mehta:2020kwu,Unal:2020jiy,Chen:2019fsq,Chen:2021lvo,Chen:2022oad,Shakeri:2022usk}, and the nature of its couplings with visible matter can be constrained by various observables  \cite{Harari:1992ea,Fedderke:2019ajk,Diego-Palazuelos:2022dsq,Luo:2023cxo,Wouters:2013hua,Marsh:2017yvc,Reynolds:2019uqt,Dessert:2019sgw,Buschmann:2019pfp,Dessert:2020lil,Dessert:2021bkv,Reynes:2021bpe,Nguyen:2023czp,Mirizzi:2009aj,Horns:2012pp,Meyer:2013pny,Fermi-LAT:2016nkz,Meyer:2016wrm,Meyer:2020vzy,Mastrototaro:2022kpt,VanTilburg:2020jvl,Lancaster:2019mde,DeRocco:2022jyq,Wadekar:2021qae, Wadekar:2022ymq,Gan:2023swl,Hook:2017psm,Balkin:2022qer,Huang:2018lxq,Caputo:2018ljp,Safdi:2018oeu,Caputo:2018vmy,Sun:2021oqp,Buen-Abad:2021qvj,Sun:2021oqp,Sun:2023gic,Chen:2019fsq,Chen:2021lvo,Chen:2022oad,Shakeri:2022usk,Graham:2023unf,Graham:2024hah}.

Though usually not thermally produced, fuzzy dark matter can still have a significant relativistic component, for instance post-inflation axions produced from relaxation of string networks. Ref.~\cite{Amin:2022nlh} pointed out that in such a relativistic regime, wave dark matter exhibits free streaming behavior much like the collisionless damping of warm or hot particle dark matter \cite{Bond:1983hb}.   Such dark matter is thus warm and fuzzy simultaneously.  

Ref.~\cite{Amin:2022nlh} highlights the apparent differences between free streaming behavior associated with the wavenumber distribution for wave dark matter and that with the particle momenta distribution for particle dark matter, and thus their respective effects on cosmological perturbations. These apparent differences are important to understand when applying free streaming considerations to bounds on the dark matter mass and the evolution of isocurvature fluctuations from post-inflation causal production as compared to cold dark matter isocurvature perturbations.

In this work, we further explore the relationship between the free streaming of wave and particle dark matter and resolve their apparent differences.  We begin in \S \ref{sec:freestreaming} by relating the particle and wave pictures of free streaming and the impact of wavenumber versus particle momentum distribution on the transfer function of density perturbations.  We show that axion wave dark matter produced after inflation is warm in this sense and only moderately enhances the Jeans or free streaming damping already present for initially cold axions.  In \S \ref{sec:sims}, we study with simulations the effect of free streaming on the causally produced isocurvature fluctuations of an even hotter, i.e.\ more relativistic, wave dark matter than axions, and resolve the paradox that the effective number density fluctuations do not damp even though the waves that compose them do -- for particles, the initial number density fluctuations are averaged out over the free streaming volume; for waves, free-streaming damping causes the momentum or wavenumber distribution to become incoherent, effectively transferring power from spatial inhomogeneities to anisotropies in the momentum distribution.   In \S \ref{sec:incoherence}, we show that the impact of the incoherence of these fluctuations prevents their appearance in time- or spatially-averaged quantities, and should be thought of as the wave analogue of multiple streams in the phase space density.  We discuss the implication of these results in \S \ref{sec:discussion} and provide appendices on the computation of free streaming (App.\ \ref{sec:lfsscaling})  for general wavenumber or momentum distributions and their impact on mass bounds (App.\ \ref{sec:averaging}).  
Throughout we employ units where $\hbar=c=1$.

\section{Free Streaming Duality}
\label{sec:freestreaming}

The free streaming of wave and particle dark matter shares the same underlying principles and can impact cosmological structure formation on scales smaller than the maximal free streaming scale.   
Given a particle mass $m$ and comoving momentum $q$, a non-interacting particle will stream with a velocity 
\begin{equation}
v= \frac{q}{\sqrt{q^2 + a^2m^2}},
\end{equation}
where $a$ is the scale factor.   Similarly, a free wavepacket of a field $\phi$ that obeys a relativistic wave equation such as the Klein-Gordon equation, with a dispersion relation $\omega = \sqrt{q^2 + a^2m^2}$, will propagate  at the group velocity 
\begin{equation}
v=\frac{\partial \omega}{\partial q} = \frac{q}{\sqrt{q^2 + a^2m^2}}
\end{equation}
around a comoving wavenumber $q$.  We use the terms momentum and wavenumber interchangeably throughout.

In both cases  the free streaming length becomes
\begin{equation}\label{eqn:lfs}
    \lfs(q;a)  = \int d\tau v(q,\tau) = \int \frac{d\ln a}{a H(a)}\frac{q}{\sqrt{q^2+a^2m^2}},
\end{equation}
where $\tau=\int dt/a$ is the conformal time.   
Where no confusion should arise, we suppress the evaluation scale factor (``$a$") in the argument of functions, e.g. $\lfs(q) \equiv \lfs(q;a)$.  

For ultrarelativistic momenta $q\gg am$, $v\approx 1$ and $\lfs(q)\approx \tau$.  For nonrelativistic momenta $q\ll am$, $v\approx q/am$, and the free streaming length grows logarithmically from its value of $\tau|_{a=q/m}$ during radiation domination and ceases to grow during matter domination.  It is therefore convenient to scale $\lfs(q)$ to the comoving Hubble length at equality $a=a_{\rm eq}$
\begin{equation}\label{eqn:lfsscaled}
\lfs(q;a) \approx  \frac{\sqrt{2}} {a_{\rm eq}H_{\rm eq}} F\left(\frac{q}{a_{\rm eq} m};\frac{a}{a_{\rm eq}}\right)\equiv \frac{\sqrt{2}} {a_{\rm eq}H_{\rm eq}} F(\hat{q};y),
\end{equation} 
where 
\begin{equation}
\frac{\sqrt{2}}{a_{\rm eq}H_{\rm eq}} =\frac{1}{H_0} \sqrt \frac{a_{\rm eq}} { \Omega_m },
\end{equation}
and carry the scaling behaviors in the various regimes with the dimensionless function $F$.   The exact analytic form of this function and its scaling behaviors are given in Appendix \ref{sec:lfsscaling}.   

Of particular interest for viable dark matter models that mimic CDM on large scales are candidates that become nonrelativistic well before equality.  The maximal scale for the impact of free streaming is the value that $\lfs$ achieves well after equality. Combining these two limits, we find the asymptotic approximation
\begin{eqnarray}\label{eqn:Flimit}
F(\qaem;y) &\approx &\qaem \ln(8/\qaem) \\
&=& \frac{ q}{a_{\rm eq} m} \ln\left( \frac{8 a_{\rm eq} m}{q}\right)
, \quad q\ll a_{\rm eq} m, a\gg a_{\rm eq}.\nonumber
\end{eqnarray}
The log term represents the logarithmic growth from the epoch that the momenta become nonrelativistic $a_{\rm nr} \sim q/m$ through $a_\eq$, and the $\hat{q}= q/a_\eq m$ prefactor likewise scales the comoving horizon at equality to $a_{\rm nr}$ given $\sqrt{2} a H(a)/a_{\rm eq} H_{\rm eq} = a_{\rm eq}/a$ in radiation domination.  

The distinction between various types of dark matter therefore mainly comes from their momentum or wavenumber distributions.
For thermally produced dark matter, this comes from the distribution at the relevant temperature for production and kinetic decoupling.  
For non-interacting scalar wave dark matter $\phi$, the number density scales as $m \phi^2$ in the nonrelativistic regime, and thus the momentum spectrum is provided by the power spectrum of $\phi$ itself. Below we will use the terms power spectrum of field fluctuations and momentum distribution of the number density interchangeably where no confusion should arise.

For the axion, if the Peccei-Quinn symmetry breaking occurs after inflation, the axion field is uncorrelated across different horizon patches, resulting in white-noise field fluctuations above the horizon. At the critical time when axions acquire masses
their potential becomes 
\begin{equation}
V(\phi) = m^2 \fax^2 [1-\cos(\phi/\fax)]
\end{equation}
and once $H(a_*)=m$, the potential energy stored in the random initial field $\phi/\fax \in [-\pi,\pi)$
will convert to locally coherent oscillations of the axion field, which produces mostly cold axions whose spatial number density varies from horizon patch to patch. This mechanism is known as vacuum misalignment production.

Furthermore, the post-inflationary axion also predicts the existence of topological defects such as axion strings due to the Kibble mechanism \cite{Kibble:1976sj}. 
The string network evolves in such a pattern that the number of strings per horizon is nearly constant, and the energy stored in string cores is lost through the radiation of relativistic axion waves \cite{Harari:1987ht,Vaquero:2018tib}.  
This emission may contribute significantly to the axion relic density \cite{Buschmann:2019icd,Buschmann:2021sdq,Gorghetto:2020qws} and extend the axion momentum distribution to $q>a_* m$, providing a ``warm" component.   
                                   
In the post-inflationary case, the power spectrum of field fluctuations then gives the momentum spectrum of the average number density of axions {\it after} the relevant momenta become non-relativistic
\begin{equation}
\langle \phi^2 \rangle = \int \frac{d^3 q}{(2\pi)^3} P_\phi(q) =\int d\ln q \frac{q^3}{2\pi^2}P_\phi(q)
\end{equation} 
with
\begin{equation}
\langle \phi(\bq)\phi(\bq') \rangle =
(2\pi)^3 \delta(\bq+\bq') P_\phi(q).
\end{equation}
This spectrum is white $P_\phi=$ const.\ for $q\ll q_* \equiv a_* m$.   For 
$q \gg q_*$, the axions produced by misalignment and by decay of the scaling string network at higher redshifts $a<a_*$ dilute their number density as $n \propto a^{-3}$ (e.g.~\cite{Sikivie:2006ni}) leading to the scaling expectation $dn/d\ln q \propto q^{-1}$ \cite{Harari:1987ht}. 
Following \cite{Amin:2022nlh}, we combine these behavior for $a \gg a_*$ as 
\begin{equation}\label{eqn:Pphi}
q^3 P_\phi(q) \propto \left( \frac{q}{q_*}\right)^3 \theta(q_*-q) + \left( \frac{q_*}{q} \right)^\alpha \theta(q-q_*) .
\end{equation}
Simulations differ on whether the $q>q_*$ power law $q^{-\alpha}$ is strictly the scaling value of $\alpha=1$ and therefore how much of the string network {\it energy} is radiated at a given momentum, which can make a large change in the overall relic number abundance \cite{Buschmann:2019icd,Buschmann:2021sdq,Gorghetto:2020qws}.  Notice however that as long as  $\alpha>0$, the number density spectrum is still dominated by momenta around $q_*$ as is the energy density $\rho \approx m n$ after all momenta are non-relativistic (cf.~Eq.~\ref{eqn:totaldensity}).

On the other hand around $q_*$, we have simply joined the two asymptotic behaviors as a broken power law spectrum.  In simulations of string dynamics, the spectrum around $q_*$ is smoother and can have transient plateau-like features before the asymptotic $q\gg q_*$ break \cite{Vaquero:2018tib,Buschmann:2019icd,OHare:2021zrq}.  In the main paper we will simply assume this broken power law form with $\alpha=1$ and in App.~\ref{sec:averaging} we explore variations and their consequences (see also \cite{Amin:2022nlh}v2, their App.~B).

Since this number density spectrum of axions is peaked around $q_*$, we can expect that the net effect of averaging the free streaming of the momenta components in the spectra is dominated by these $\sim q_*$ modes, which are only quasirelativistic or ``warm" at birth.   Correspondingly, we would expect the impact of free streaming on density perturbations to occur at (see App.~\ref{sec:averaging} and Fig. \ref{fig:comparison_qfs})
\begin{eqnarray}\label{eq:kfs_def}
k_{\rm fs} &\equiv&  \lambda_{\rm fs}^{-1}(q_*=a_*m) 
= \frac{\sqrt{a_{\rm eq}^2 m H_{\rm eq}}}{2^{1/4}\ln ( 8a_{\rm eq}/a_* )}.
\end{eqnarray}
Throughout, when we compute numerical values for such quantities we take a cosmology with matter density
$\Omega_m h^2 = 0.142$ and 3 massless neutrinos. 

This scale should be compared to the similar suppression of the transfer function at the Jeans scale in the case of  Pecci-Quinn symmetry breaking before the end of inflation.  Here the initial misalignment is coherent across the whole horizon volume today by the end of inflation and there is only an initially cold component to the axions.   Curvature fluctuations of wavenumber $k$ then imprint field fluctuations $\phi(q)$ at wavenumber $q=k$ and the density fluctuations are carried by $\phi^2(k) \approx 2 \phi(k) \langle \phi\rangle$.
Throughout, ``$k$" identifies the wavenumber of quadratic quantities such as $\phi^2$, number, and energy density, whereas ``$q$" when different from $k$ distinguishes the wavenumber of field fluctuations that compose them.

\begin{figure}
    \centering
    \includegraphics[width=1.0\linewidth]{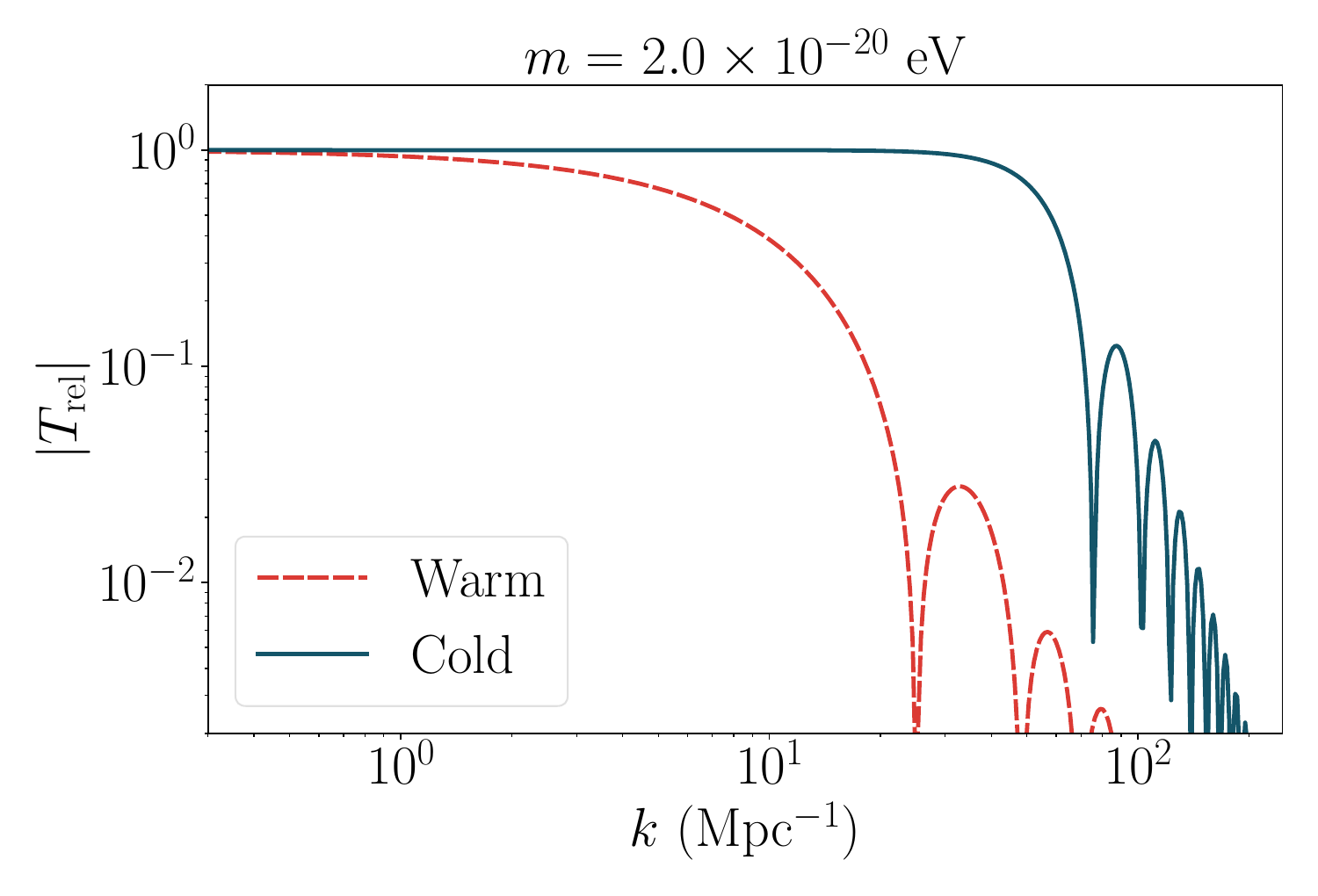}
    \caption{Relative transfer function (\ref{eqn:transfer}) due to the effect of free-streaming on the ``warm" axions of the post-inflationary mechanism compared with that of the cold axions of the pre-inflationary case. 
    Here  $m = 2.0\times 10^{-20}$ eV and the warm spectrum peaks at $q_* = a_* m$, the horizon wavenumber at the start of axion oscillations, and the evaluation epoch $a=0.2$ is chosen to reflect that of the Lyman-$\alpha$ forest. 
}
\label{fig:transfer}
\end{figure}

The relevant free streaming scale for the pre-inflationary case is the comoving wavenumber whose associated free streaming length overtakes its wavelength 
\begin{equation}\label{eqn:fstoJeans}
\lfs(k_J) \approx \lambda_J\approx \sqrt{6}/k_J,
\end{equation}
and by 
employing Eq.~(\ref{eqn:Flimit}), we have
\begin{equation}\label{eqn:Jeans}
k_J \approx 3^{1/4}  \sqrt{ \frac{ a_{\rm eq}^2 m H_{\rm eq}}{\ln ( 8a_{\rm eq}/a_* )}},
\end{equation} 
where we have ignored self-interactions \cite{Khlopov:1985fch}.
Note that the $\sqrt{6}$ in Eq.~(\ref{eqn:fstoJeans}) is added so that Eq.\ (\ref{eqn:Jeans}) matches the definition in the literature \cite{Hu:2000ke} (their Eq.~9), modulo the log factor which we have here approximated at $k_J \sim  a_* m = 2^{-1/4} a_{\rm eq} \sqrt{ m H_{\rm eq}}$ but is usually incorporated more precisely as a mass-dependent fitting factor to numerical calculations of the transfer function (\cite{Passaglia:2022bcr} their Eq. 44).

We therefore expect that the free streaming of the quasirelativistic or ``warm" axions in the post-inflationary scenario to scale in the same way as the Jeans scale of cold axions in the pre-inflationary scenario and differ only by the ratio
\begin{equation}\label{eqn:kratio}
\frac{k_J}{k_{\rm fs}} = {6^{1/4}} 
\sqrt{ \ln ( 8a_{\rm eq}/a_* )}.
\end{equation}
We can improve on this estimate by averaging over the momentum spectrum $q^3 P_\phi$ instead of evaluating at the peak to define an effective transfer function for density perturbations due to the free streaming effect following Ref.~\cite{Amin:2022nlh}
\begin{equation}\label{eqn:transfer}
T_{\rm rel}(k)\equiv
\frac{T_{\rm ax}}{T_{\rm CDM}}(k) \approx \frac{\int d\ln q \frac{q^3}{2\pi^2}  P_\phi(q) \frac{\sin[k\lfs(q)]}{k\lfs(q)}}{\int d\ln q \frac{q^3 }{2\pi^2}P_\phi(q)},
\end{equation}
 who derive this form from the WKB approximation for the amplitude of the free-streaming waves and linearizing adiabatic perturbations as a means of estimating where free streaming has an $O(1)$ effect.  Here $T_{\rm CDM}$ is the cold dark matter (CDM) density transfer function (with no axions) and $T_{\rm ax}$ is the axion transfer function (with no CDM),  with the same cosmological parameters otherwise. 
 This ratio serves to isolate the free streaming effect by removing other cosmological effects from matter radiation equality and baryon acoustic oscillations.
The amplitude reduction term can also be motivated from the treatment of wave propagation in \S \ref{sec:sims} where free streaming modifies the initial wave amplitude for each momentum $q$ according to the solution to the wave equation
(see Eq.~(\ref{eqn:relativisticphi})).
We will discuss the impact of the slowly varying phases of the momentum components in \S \ref{sec:incoherence}.
\begin{figure}
    \centering    \includegraphics[width=\linewidth]{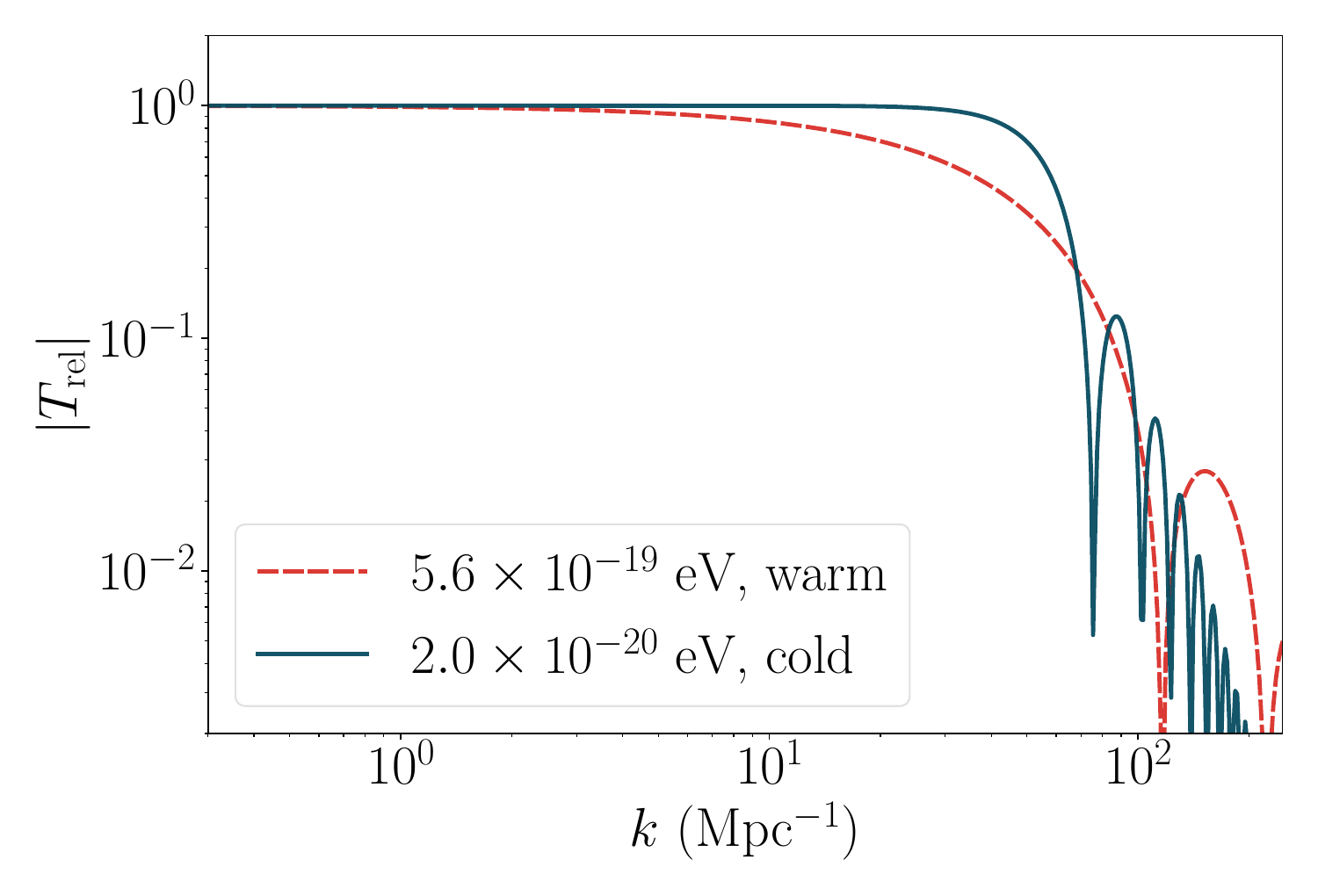}
    \caption{
    Mass scaling of the relative transfer function of warm axions relative to the cold axions in Fig.~\ref{fig:transfer}.  As predicted from Eqn.~(\ref{eqn:kratio}), a warm axion mass of $5.6\times10^{-19}$ eV produces a comparable scale of suppression to cold axions of $2.0\times 10^{-20}$ eV.
    }
    \label{fig:transfer_with_shiftedmass}
\end{figure}
In Fig.~\ref{fig:transfer}, we compare the relative transfer function (\ref{eqn:transfer}) with the usual Jeans suppression for cold axions from a numerical calculation using \textsc{AxiECAMB}, a modified version of \textsc{CAMB}\footnote{\textsc{CAMB}: \url{http://camb.info}\\
\textsc{AxiECAMB}: \url{https://github.com/Ra-yne/AxiECAMB}}\cite{Liu:2024yne} (see also \cite{Passaglia:2022bcr} their Eq.~44, which closely match these results).  We illustrate this with a mass of $m=2.0\times 10^{-20}$eV which is motivated by the bound on cold axions from the Lyman$-\alpha$ forest \cite{Rogers:2020ltq}. Correspondingly we take a redshift of $z=4$ or $a=0.2$ as the evaluation epoch.  As expected, the free streaming transfer function for the ``warm" axions gives a stronger suppression than the cold, pre-inflationary case but only by a log factor.  In fact, Eq.~(\ref{eqn:kratio}) predicts $k_J/k_{\rm fs}\approx 5.3$, which captures most of the difference. This ratio can be then be used to approximately scale up any given Lyman-$\alpha$ bound on cold axions since $k_J\propto  m^{1/2}$, here nominally $m \gtrsim 5.6 \times 10^{-19}$eV. Fig.~\ref{fig:transfer_with_shiftedmass} demonstrates that warm axions of this mass give a transfer function comparable to the ``cold'' axions of $m = 2.0\times 10^{-20}$ eV.

For heavier masses, where free streaming is negligible on observationally relevant scales, the random number of cold axions in each horizon patch at $a_*$ leads to so-called isocurvature fluctuations on large scales, which is also constrained by the Lyman-$\alpha$ forest observations. 
At large scales, the isocurvature perturbation is well described by the white-noise power spectrum and is not sensitive to the behavior of the power spectrum at $k \sim q_*$ which determines the free-streaming effect. Therefore, the ratio of the amplitude of the isocurvature fluctuations to that of the adiabatic is $f_{\rm iso}^2\propto 1/q_*^3$. 
A lower $q_*$, corresponding to a lighter axion in this range, gives a larger isocurvature to adiabatic ratio on large scales, imposing a lower bound on the axion mass in the post-inflationary scenario, $m>3\times 10^{-18}$ eV from Lyman-$\alpha$ forest constraints  \cite{Irsic:2019iff}, which is stronger than the nominal free streaming bound above.
For this mass, the free-streaming length $\lfs(q_*) = 0.014$\,Mpc, which implies that the free streaming effect can be neglected when applying this isocurvature bound.

\section{Causally Coherent Patches}
\label{sec:sims}

We have seen in the previous section that for axion dark matter produced around the beginning of the oscillation epoch $H(a_*)=m$ with only a mildly relativistic or ``warm" component, the free-streaming scale for the post-inflationary production mechanism remains close to that of cold axions from the pre-inflationary production mechanism.  Thus, the main new large scale effect in the post-inflationary case is that the causally random patches on the horizon scale at $a_*$ produce a white noise spectrum of axion number or mass density fluctuations.

For a hypothetical wave dark matter candidate produced causally after inflation with more relativistic momenta, the free streaming scale can be larger; simultaneously, the white noise fluctuations can be smaller in amplitude on a fixed observational scale given the smaller horizon scale at production, and therefore weaken the isocurvature bounds on the mass until the free streaming limit comes to dominate \cite{Amin:2022nlh}. 

More concretely as shown in App.~\ref{sec:averaging}, if the number density spectrum peaks at $q_{\rm peak}= R q_* = R a_* m$, then the effective free streaming length increases as $\lfs(q_{\rm peak}) \propto R$ up to a log correction using Eq.~(\ref{eqn:Flimit}).  For moderate increases where $R \sim {\cal O}(1)$, this can make free streaming more important and become competitive or stronger than the isocurvature bound.
The combination of the two is therefore more robust for changes in the axion spectrum for masses in the $m \sim 10^{-18}-10^{-19}$eV range \cite{Amin:2022nlh}.

For $R\gg 1$,
most of the scalar wave dark matter is ultrarelativistic at $a_*$.  In this case, the dark matter is ``hot" at birth.  Note that this does not occur with axions, since at any given epoch after Peccei-Quinn symmetry breaking but before $H\sim m$, the Kibble mechanism establishes a coherent field with a random value of $\phi/\fax \in [-\pi,\pi)$ across the Hubble patch $1/aH$ with a self similar network of strings and their decay products.

In the more general case, it is also important to understand the evolution of the isocurvature density fluctuations from the causally random initial field fluctuations, since free streaming suppression and isocurvature enhancement work in opposite directions.  In Ref.~\cite{Amin:2022nlh}, it was shown that a certain characterization of these field fluctuations, namely fractional fluctuations in $\phi^2$, remains constant and white in power despite the free streaming of the waves that compose them.  In this section, we seek to clarify the nature of these fluctuations from the standpoint of the free streaming of the causally coherent initial horizon scale patches.

To understand the impact of free-streaming on causally coherent patches, let us begin with an initial field profile that is a spherical tophat\footnote{We have also tested Gaussian profiles and found similar results. We choose tophat here for clarity of wavefront visualizations.}
of radius $\tau_i$ and set the field normalization to unity  at $r<R$.   The Fourier transform of the tophat gives the momentum distribution of the patch 
\begin{equation}\label{eqn:tophatq}
\phip(q;0) = \frac{3 V_p}{(q\tau_i)^3}\left[{\sin(q \tau_i) - q \tau_i \cos(q \tau_i)}\right],
\end{equation}
where $V_p =4\pi \tau_i^3/3$ is the volume of the tophat patch.
The spectrum is a constant $\phip(q;0) = V_p$ for   $q\tau_i \ll 1$ and a random distribution of these causal patches would produce white noise on large scales as desired.

\begin{figure*}
    \centering
       \includegraphics[width=\linewidth]{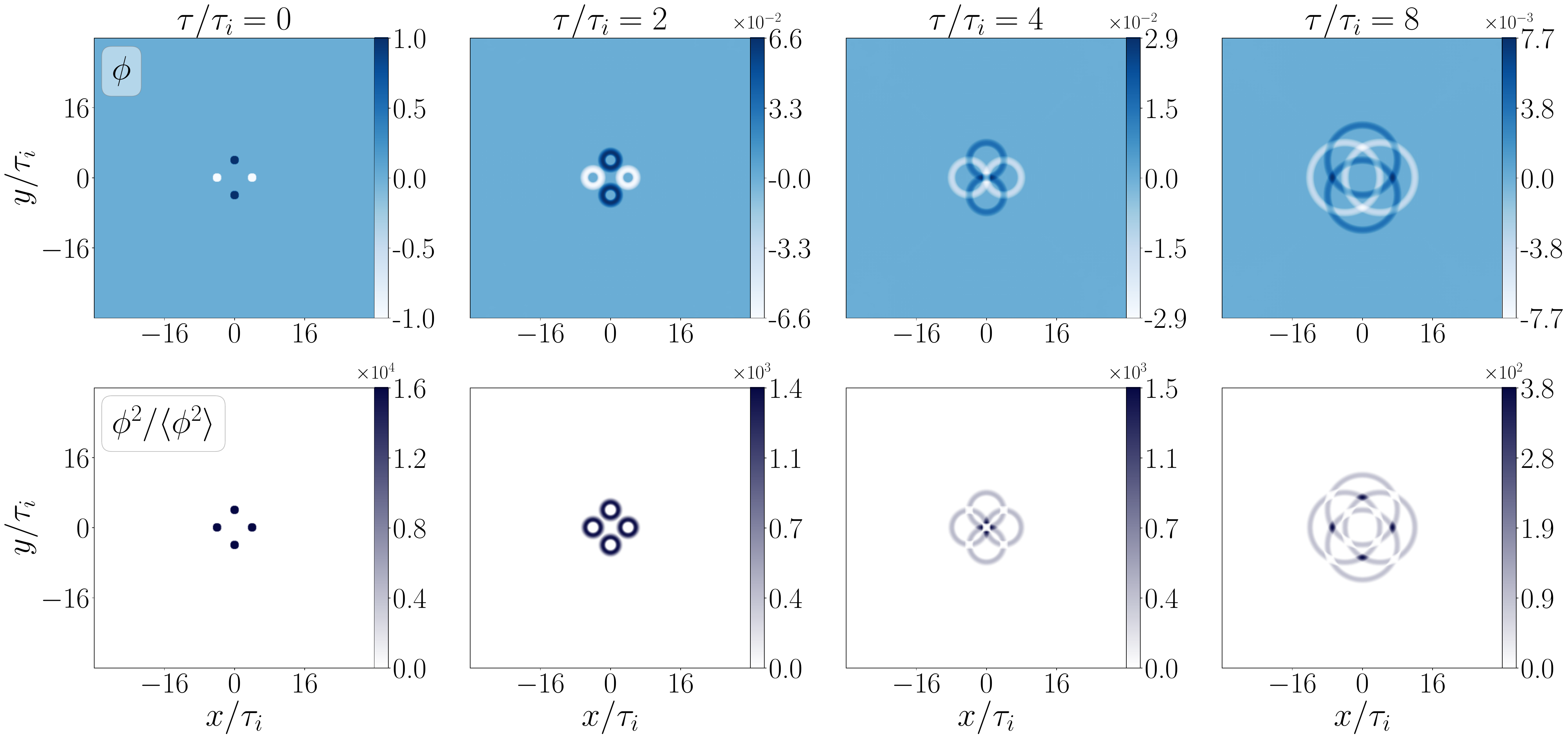}
    \caption{
    Free streaming of 4 coherent tophat field patches (see text) 
    for the field $\phi$ (top panel) and the proxy for number density fluctuations $\phi^2/\langle \phi^2\rangle$ (bottom panel) for a series of snapshots in conformal time $\tau$.  As the waves from different patches free stream, they individually damp in amplitude and spread in scale.  When multiple wave fronts intersect and superimpose, fluctuations appear that are transient and reflect multiple momentum streams from different patches.}    \label{fig:4patches_vis}
\end{figure*}

\begin{figure}
    \centering
    \includegraphics[width=1\linewidth]{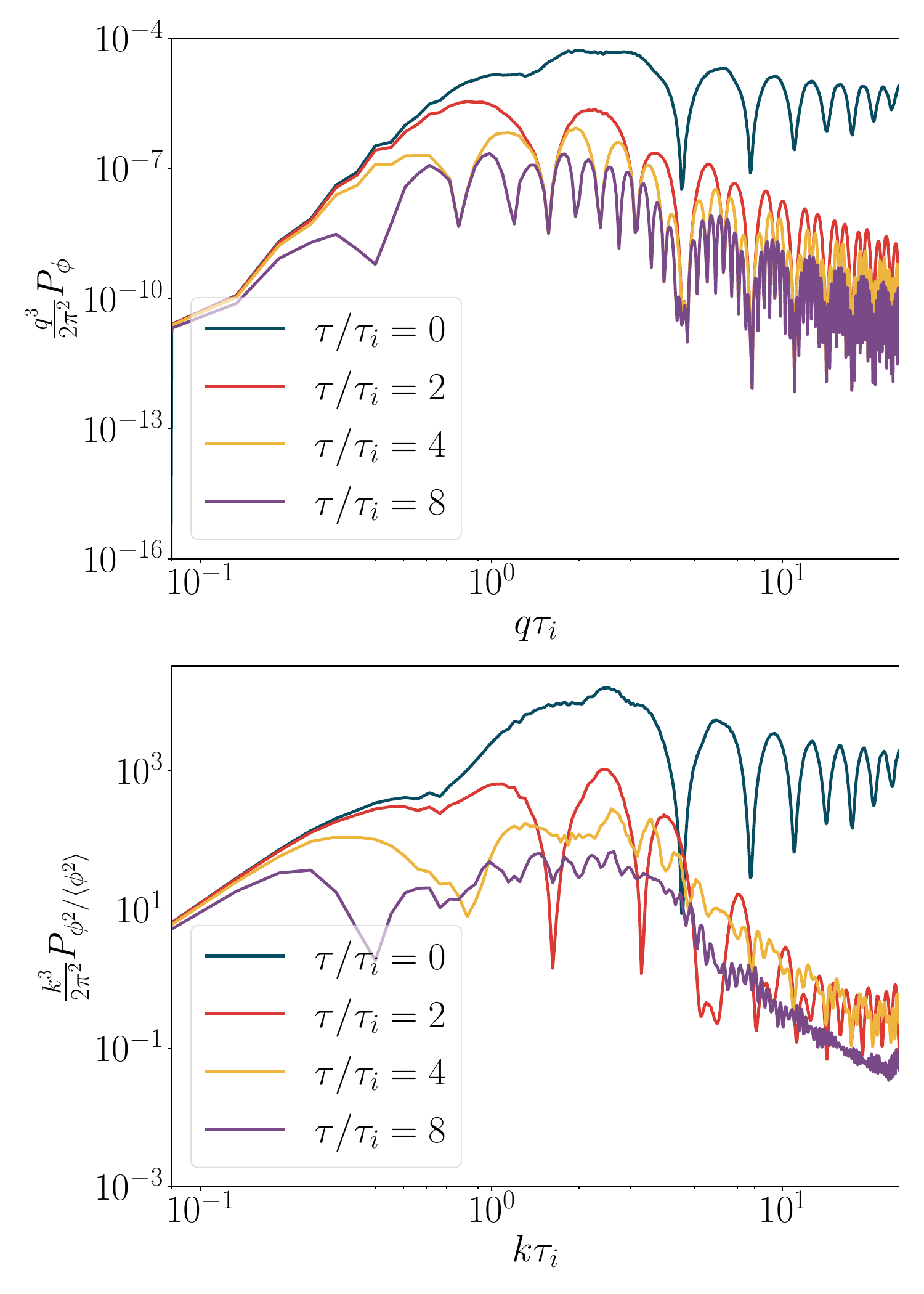}
    \caption{Power spectra of $\phi$ and $\phi^2/\langle \phi^2 \rangle$ for the 4 patch 
case of Fig.~\ref{fig:4patches_vis}.  Free streaming damping in both cases scales as $k \propto \lfs^{-1} =\tau^{-1}$ and here the phase-coherent autocorrelation of momentum modes from individual patches dominates.
    }
    \label{fig:4power}
\end{figure}

The Klein-Gordon equation in $q$-space for a free field 
\begin{equation}\label{eqn:KG}
\ddot \phi + 2a H \dot\phi + (q^2 + a^2 m^2)\phi = 0,
\end{equation}
evolves the initial modes,
where overdots denote conformal time derivatives. More specifically, each mode evolves via the growth function
\begin{equation}\label{eqn:phievolution}
\phi(q;\tau) = \phi(q;0) D(q;\tau),
\end{equation}
where $D$ solves the Klein-Gordon equation (\ref{eqn:KG}) with an initially frozen field due to the Hubble drag,
$\dot\phi(q;0)=0$.
During radiation domination,  $\tau=1/aH$ and $a^2m^2 = \tau^2/\tau_*^4$, and 
the growth function of the field is given by
\begin{equation}\label{eqn:generalD}
D(q;\tau)  =  e^{-i (\tau/\tau_*)^2/2} {}_1 F_{1} \left[\frac{3}{4} + i \frac{(q\tau_*)^2}{4}, \frac{3}{2}, i \left(\frac{\tau}{\tau_*}\right)^2\right],
\end{equation}
where ${}_p F_q$ is the generalized hypergeometric function. For $\tau\ll \tau_*$, $D(q;\tau)$ takes the simple form
\begin{equation}\label{eqn:relativisticphi}
\lim_{\tau/\tau_* \ll 1} D(q;\tau) = \frac{\sin (q\tau)}{q\tau} ,
\end{equation}
(cf.~Eq.~\ref{eqn:transfer}).
Here the field is frozen above the horizon $q\tau\ll 1$ and oscillates with amplitude $D \propto 1/\tau \propto 1/a$ below the horizon.   This amplitude decay reflects the redshifting of relativistic waves inside the horizon.  Furthermore, the number density associated with $q$ goes as $n \propto (\omega/a) \phi^2 \propto a^{-3}$, and particle number in each mode is conserved in comoving coordinates.

With the initial tophat wavepacket, Fourier transforming the product of Eq.\ (\ref{eqn:tophatq}) and
(\ref{eqn:relativisticphi}) gives
in real space
\begin{equation}
\phip(r;\tau) = 
\begin{cases}
			1, &  r<\tau_i-\tau, \tau<\tau_i
			 \\
			 			0, &  r<\tau-\tau_i, \tau\ge \tau_i
			 \\
           \frac{(\tau_i+\tau -r)(\tau_i-\tau+r)}{4 r\tau}, &  | \tau_i-\tau |\le r < \tau_i+\tau \\
           0, & r>\tau_i+\tau
\end{cases}
\label{eqn:tophatevol}
\end{equation}
which represents a spherically symmetric shell expanding radially where $r=|\bx |$ with a front at $\tau_i+\tau$, reflecting causal propagation at  the speed of light.  
As expected, the field amplitude damps as it spreads outwards and transfers its coherent fluctuations to the larger physical scales associated with the free streaming scale $\lfs =\tau$. 

For $\phi_p^2$, its Fourier components are composed by a convolution of the field momenta
\begin{eqnarray}
\phip^2(\bk;\tau) &=& \int \frac{d^3 q}{(2\pi)^3} 
\phip({\bq};\tau) \phip({\bk-\bq};\tau) \nonumber\\
&=& \int \frac{d^3 q}{(2\pi)^3} 
\frac{\sin(q\tau)}{q \tau} \frac{\sin(|\bk-\bq|\tau)}{|\bk-\bq|\tau}\nonumber\\
&&\times 
\phip({\bq};0) \phip({\bk-\bq};0).
\end{eqnarray} 
Since the initial profile is coherent at $r<\tau_i$, different field momenta $\bf q$ are correlated in their phase and coherently superimpose in this quadratic combination to produce the spatially coherent fluctuation $\phi_p^2(\bx;\tau)$.
As with $\phi_p$, the power spectrum of $\phi_p^2$ is strongly damped by free streaming and represents the dilution or averaging out of the coherent fluctuation in a given patch. Interpreted in the particle picture, the initial axion number fluctuation in a given coherent patch is averaged out over the free streaming scale.

On the other hand, the total $\phi$ is a sum over all horizon patches, each with a random amplitude, and $\phi^2$ receives contributions not just from the coherent propagation of modes of a single patch but also all of the phase-incoherent cross terms between patches.  In this case, there is a sum over $N$ patches: 
\begin{eqnarray}
\phi(\bq;\tau) & =& 
\sum_{\alpha=1}^N \phi_\alpha(\bq;\tau), \nonumber\\
\phi_\alpha(\bq;\tau) &\equiv &
 A_\alpha \phi_p(\bq;\tau) e^{i \bq \cdot {\bf d}_\alpha},
\end{eqnarray}
where $A_\alpha$ is the field value at the center of patch $\alpha$ at spatial coordinate ${\bf d}_\alpha$. Correspondingly,
\begin{eqnarray}\label{eqn:phi2}
\phi^2(\bk;\tau) & = &\sum_{\alpha\beta}
\phi_{\alpha\beta}^2(\bk;\tau) ,\nonumber\\ \phi_{\alpha\beta}^2(\bk;\tau) &\equiv& 
\int \frac{d^3 q}{(2\pi)^3} 
\frac{\sin(q\tau)}{q \tau} \frac{\sin(|\bk-\bq|\tau)}{|\bk-\bq|}\nonumber\\
&&\times  
\phi_\alpha({\bq};\tau) \phi_\beta({\bk-\bq};\tau),
\end{eqnarray}
 now has autocorrelation terms between modes from the same patch $\alpha=\beta$ and cross terms between patches $\alpha\ne \beta$.   Both the evolution and the physical interpretation of the auto and cross terms differ.  In particular, for the autocorrelation terms, the power spectrum of $\phi^2_{\alpha\alpha}= \phi^2_p$ contains phase correlations between differing momenta and damps with free streaming, whereas the cross terms carry incoherent phase shifts due to the displacements ${\bf d}_\alpha$, though these vanish when pairing the same momenta in the $\phi^2$ power spectrum, i.e. $\phi(\bq)\phi(-\bq)$.

\begin{figure*}
    \centering
    \includegraphics[width=1\linewidth]{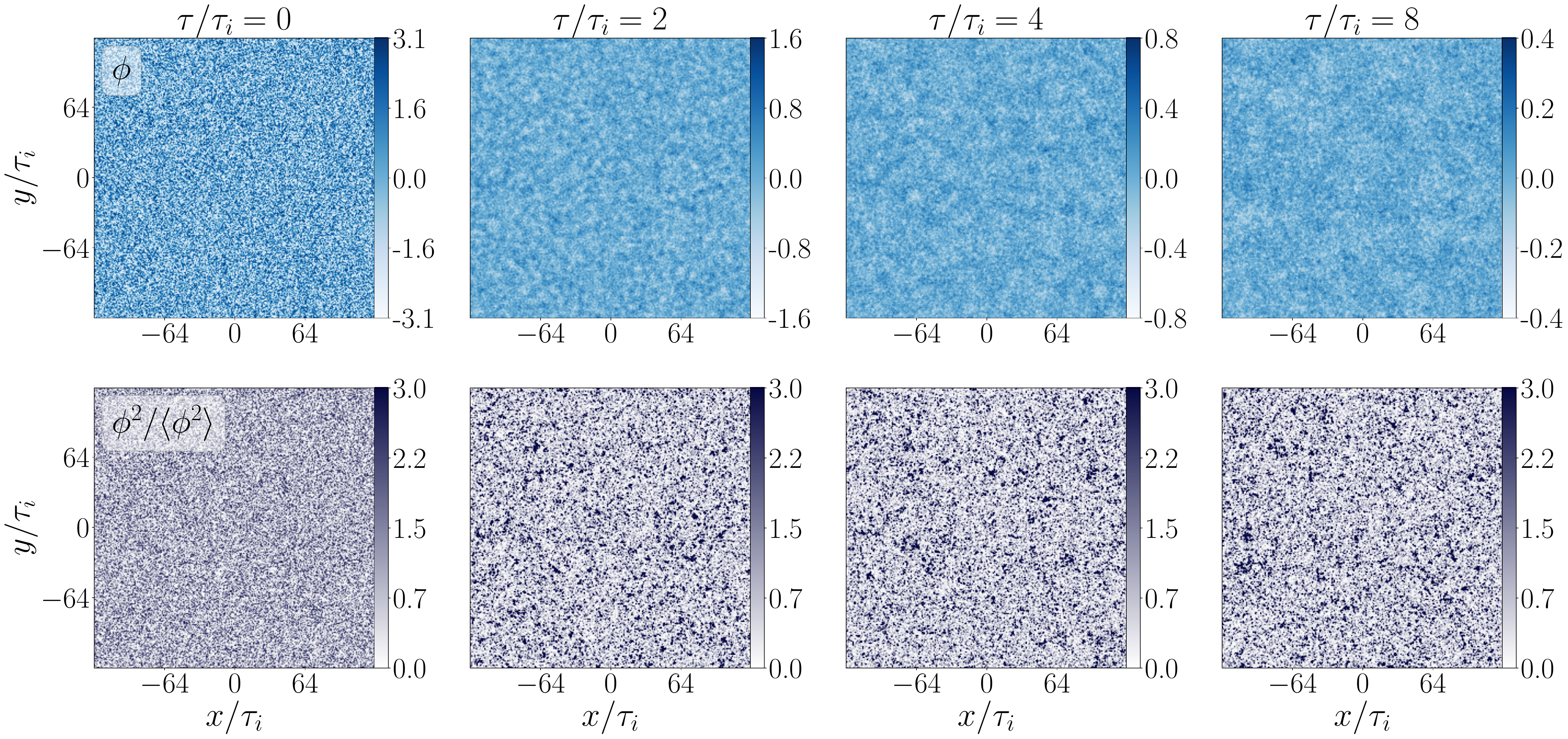}
\caption{Free streaming evolution as in Fig.~\ref{fig:4patches_vis} but with pixel scale horizon patches 
and field values drawn from a uniform distribution.  While the field $\phi$ free streams and becomes smoother and lower in amplitude with time, $\phi^2/\langle \phi^2 \rangle$ becomes statistically time invariant with transient fluctuations at a given position. 
\label{fig:uniform_random_vis}}
\end{figure*}

To visualize the difference, consider explicitly a simple superposition of such patches. In Fig.\ \ref{fig:4patches_vis} we set up  4 patches in a periodic box of length $64\tau_i$ on each side, with displacements $d_\alpha$ of $\pm 4\tau_i$ from the center of the box in the $x$ and $y$ directions, and positive and negative amplitudes respectively such that the total field has zero mean. The box is represented by $512^3$ discrete pixels, with $8$ pixels per unit $\tau_i$.  We evolve this configuration until $\tau = 8\tau_i$, again by employing Eq.~(\ref{eqn:relativisticphi}) in Fourier space. We have tested the simulation procedure by verifying that the time evolution of one such patch is consistent with the analytic radial solution (\ref{eqn:tophatevol}).  

In Fig.\ \ref{fig:4patches_vis} (top panels) we show the field profile itself in a $z=0$ slice of the box.
Notice that the free streaming of the individual patches brings the wavefronts to intersect at the center of the box at around $\tau=4\tau_i$.  Therefore the momentum distribution of the field fluctuations at the center is anisotropic, specifically quadrupolar.  More generally, at any given time after the free-streaming intersection of fronts, the total field represents a transient superposition of waves composed of multiple field momentum streams at any given physical position.  This is the same behavior as the particle free streaming of CMB photons after recombination or cosmic neutrinos after their decoupling: the initial particle number inhomogeneity becomes a phase space anisotropy after free streaming.  In those cases, the total power in fluctuations of a given $k$-mode is conserved (e.g.\ \cite{Hu:1994jd}, their Eq.~10), but its nature and effect on gravitational structure formation differ qualitatively.  In the free-streaming damping context, this is known as the directional damping of collisionless components \cite{Bond:1983hb}.

These free streaming considerations apply to $\phi^2$ as well.   In Fig.~\ref{fig:4patches_vis} (bottom panels), we show $\phi^2$ normalized to its average in the box $\langle \phi^2 \rangle$.  This normalization removes the redshifting effect of the subhorizon modes and we can see the remaining strong effect of free streaming damping of the amplitude of $\phi^2/\langle \phi^2 \rangle$ of each patch, from the change in scale of the panels with $\tau$. 
Moreover, even though $\phi^2(\bx)$ itself is not a directional quantity, its local value reflects the directionally dependent propagation of the fronts of each of the 4 patches.

\begin{figure}
    \centering
    \includegraphics[width=1\linewidth]{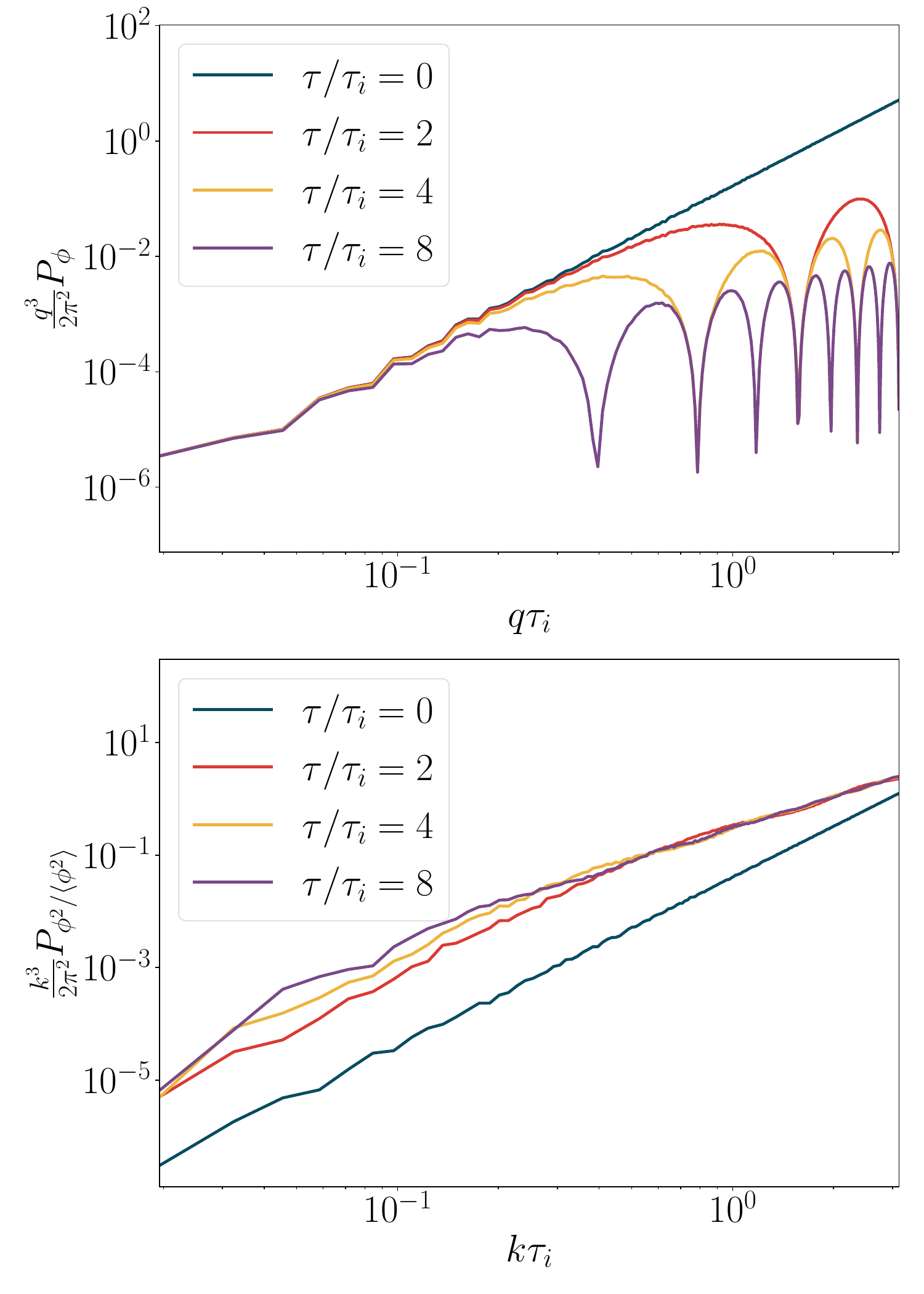}
    \caption{Power spectra of the pixel scale horizon patches of Fig.~\ref{fig:uniform_random_vis} as in Fig.~\ref{fig:4power}.  While the field power (top panel) continues to evolve and oscillate with free streaming, the power in $\phi^2/\langle \phi^2\rangle$ (bottom panel) approaches a constant power law behavior below the free streaming scale.}
    \label{fig:uniform_random_power}
\end{figure}

The corresponding power spectra for $\phi$ and $\phi^2/\langle \phi^2 \rangle$ are shown in Fig.~\ref{fig:4power}.  Note that
$P_{\phi^2/\langle \phi^2\rangle} = P_{\phi^2} / \langle \phi^2\rangle^2$.  In this 4-patch case, the power spectra themselves are still dominated by the autocorrelation terms of each patch, and the total $\phi^2$ still strongly damps with free streaming.  
We can see that the turnover into the free streaming oscillation behavior scales with the free streaming length $\lfs =\tau$,  $k \propto 1/\tau$, as expected.

On the other hand, as the number of patches $N$ grows, the number of cross terms grows as $N^2$. 
In fact, 
since the number of patches that fill a volume $V$ will be $N=V/V_p$, it is the cross terms that become the dominant source of fluctuations in $\phi^2$.  At $\tau=0$ the auto and cross correlation contributions to the power spectrum of $\phi^2$ are comparable, but the autocorrelation terms free stream away at later times.   These remaining contributions represent the incoherent superposition of fluctuations of different momenta ${\bf q}$.

From Eq.~(\ref{eqn:phi2}), we can see that cross terms have no time averaged effect on $\phi^2$ since $q$ and $|\bk-\bq|$ modes oscillate incoherently, but  provide a source of instantaneous  power 
\begin{eqnarray} \label{eqn:phi2random}
P_{\phi^2}(k;\tau) &\approx&  2 \int\frac{d^3 q}{(2\pi)^3} 
\left[\frac{\sin(q\tau)}{q \tau} \frac{\sin(|\bk-\bq|\tau)}{|\bk-\bq|}\right]^2
\nonumber\\
&&\times 
P_\phi({\bq};0) P_\phi({\bk-\bq};0),
\end{eqnarray} 
since the square of the oscillating growth function is positive definite.  Here we have dropped the autocorrelation terms that contain the phase coherence between the momentum modes, i.e.~the connected pieces of the trispectrum of $\phi$.

This behavior can be seen directly in Fig.~\ref{fig:uniform_random_vis} where we take $\tau_i$ to be the grid spacing such that each pixel represents a horizon patch.   The box size is $\Npix\tau_i$ in this simulation and the field value at each pixel is a uniform random deviate
with $\phi \in [-\pi,\pi)$.\footnote{The $[M]$  scale of $\phi$, e.g. $\fax$ for axions, drops out of the normalized quantities we consider here.} Beyond the axion context, we have also separately checked that a Gaussian random deviate produces qualitatively the same free streaming effects we describe below but with a larger number of rare high density peaks at the pixel scale.
  Again the top panels in Fig.\ \ref{fig:uniform_random_vis} show the time evolution of $\phi$ and the bottom panels  that of $\phi^2/\langle \phi^2 \rangle$.  
  Instead of the coherent propagation of discrete horizon scale patches, we are now dominated at late times by the cross terms between the $\Npix^3$ initial pixel scale patches.  
  
  Notice also that the statistical properties of the $\phi^2(\bx)/\langle \phi^2 \rangle$ field become nearly time-invariant. 
  This is in contrast to the 4 patch case even though both cases show the field $\phi(\bx)$ continuing to evolve, especially in amplitude, as their respective patches free stream.

We quantify this in Fig.~\ref{fig:uniform_random_power} for the respective power spectra.   The initially white $q^3 P_\phi(q)\propto q^3$ turns over to oscillate with a $\propto q^1$ scaling and decreasing amplitude which reflects the redshifting behavior of the growth function $D$ in Eq.~(\ref{eqn:relativisticphi}) as with the 4 patch case.  
On the other hand, the power spectrum of $\phi^2$ gains a non-oscillating and nearly constant in time $k^3 P_{\phi^2/\langle \phi^2 \rangle} \propto k^2$ spectrum on scales below the free streaming scale.  This reflects the scaling of  Eq.~(\ref{eqn:phi2random}) after normalization by $\langle \phi^2 \rangle$ which removes the redshifting effect.   

\begin{figure}
    \centering
    \includegraphics[width=1\linewidth]{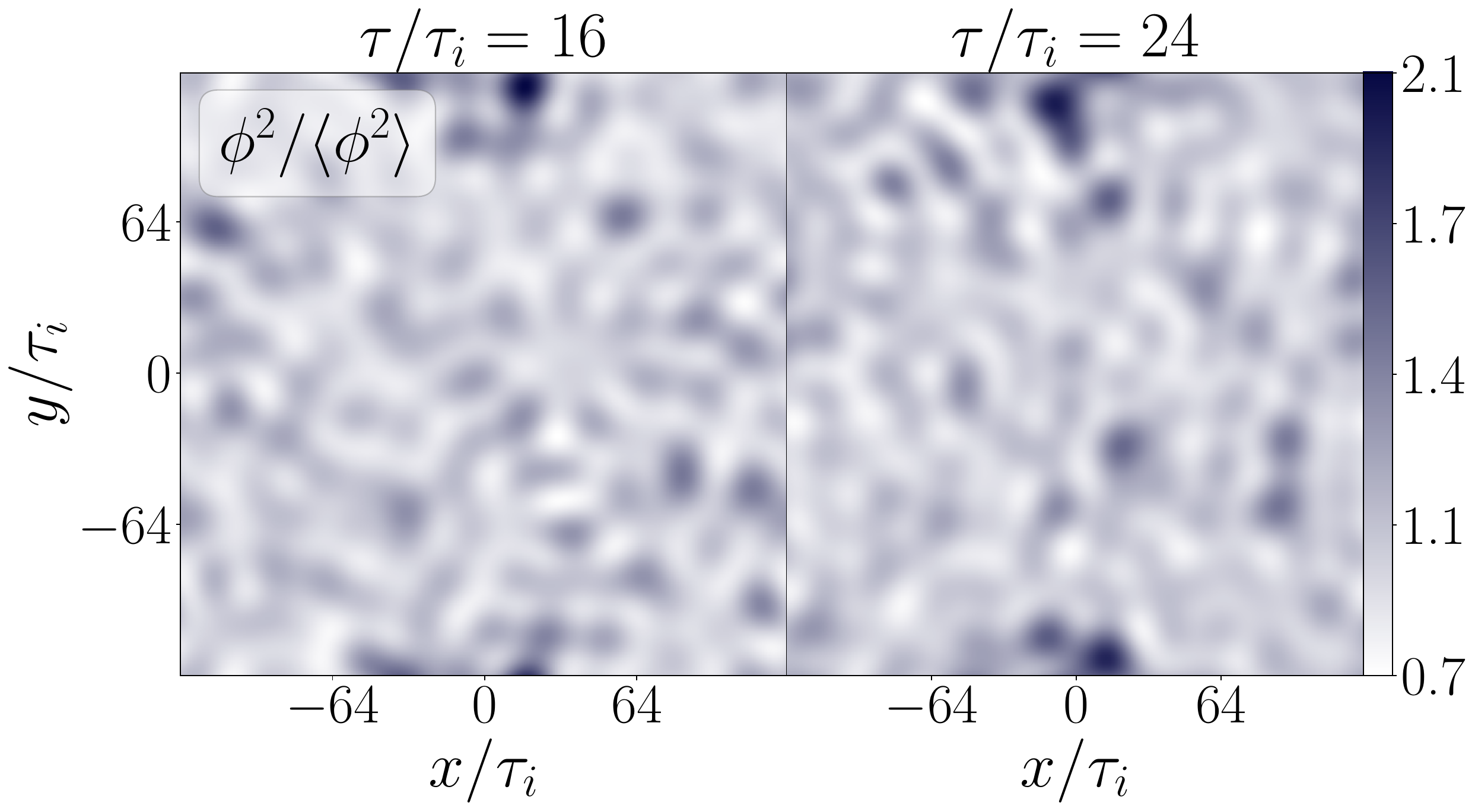}
    \caption{Long wavelength fluctuations in 
    $\phi^2$ at late times as in 
    Fig.~\ref{fig:uniform_random_vis} but low pass filtered to $k\tau_i < 0.3$.  Although statistically the same, the two snapshots at $\tau=16\tau_i$ and $24\tau_i$ show strong time evolution that reflects the transient nature of incoherent superposition.}
    \label{fig:uniform_low_pass}
\end{figure}

In particular, $\phi^2$ retains fluctuations across a range of scales below the free streaming scale but above the initial horizon scale $\tau_i$.   To better see this, we low pass filter $\phi^2$ to retain only $k\tau_i < 0.3$ and show two late time snapshots $\tau=16\tau_i$ and $24\tau_i$ in Fig.\ \ref{fig:uniform_low_pass}. Notice that even though the power in fluctuations on these scales is nearly constant, the $\phi^2$ field evolves on a time scale much shorter than the Hubble time reflecting the chance superposition of the underlying free-streaming modes.    Again, in the particle picture, this reflects a phase space anisotropy in the distribution rather than a physical space inhomogeneity.   Also, it is important to note that a low pass filter in $k$ for $\phi^2$ is not in general the same as a low pass filter in $q$ in $\phi$ since high momenta modes can still contribute to low wavenumbers $\bk = \bq_1+\bq_2$ if $\bq_1\approx -\bq_2$, which as we shall see below is the physically relevant case after all populated modes have become nonrelativistic.

In fact, during the simulated epochs where the $q$-modes are still ultrarelativistic, this free streaming behavior fully parallels that of relativistic particles and the remaining phase space fluctuations would not contribute to the gravitational formation of structure.

The distinction between wave dark matter and the relativistic particle case is that $\phi$ is a massive field and even for initially relativistic $q$ modes, the redshifting due to cosmic expansion will eventually make the modes nonrelativistic, much like the massive neutrino component of dark matter in $\Lambda$CDM.   For viable dark matter models, this occurs well before matter radiation equality and we must consider the impact of their further evolution until non-relativistic.   Since these modes are ultrarelativistic at $\tau_*$ where $H=m$ by construction, this means that we must consider their evolution at $\tau>\tau_*$ and account for the change in the spectrum as progressively larger $q$ modes become nonrelativistic.

For $\tau>\tau_*$, even the 
superhorizon modes that are non-relativistic at  $\tau_*$  evolve with $D(q;\tau)$ reflecting the coherent oscillations of the axion field due to the mass term.  
For modes that are above the horizon at $\tau_*$, $q\tau_* \ll 1$, the growth function becomes
\begin{equation}
\lim_{q\tau_* \ll 1} D(q;\tau) = \sqrt{\frac{2 \tau_*}{\tau}}\Gamma\left(\frac{5}{4}\right) J_{1/4}(\tau^2/2\tau_*^2).
\end{equation}
Note that $(\tau/\tau_*)^2/2 = mt$ where $t$ is coordinate time so that the Bessel function carries the mass scale oscillations for $mt = m/2H \gg 1$.  The amplitude of these oscillations scale as $D \propto \tau^{-3/2} \propto a^{-3/2}$.   This behavior again reflects the redshifting of non-relativistic matter $n \propto (\omega/a)\phi^2 \propto m\phi^2 \propto a^{-3}$.   Notice that for modes that remain relativistic until after  $\tau>\tau_*$, even though their number density $n \propto a^{-3}$, the extra redshifting of the frequency means that the field fluctuations decay less quickly as $a^{-1}$ instead of $a^{-3/2}$.  

In general, 
the $k$-modes of $\phi^2$ are then constructed out of the $q$-modes of $\phi$ as
\begin{eqnarray}
\phi^2(\bk;\tau) 
&=& \int \frac{d^3 q}{(2\pi)^3} D(q;\tau) D(|\bk-\bq|;\tau) \nonumber\\
&&\times 
\phi({\bq};0) \phi({\bk-\bq};0),
\end{eqnarray} 
and the different behavior of $D$ as a function of momentum $q$ changes the weights of the field fluctuations that factor into a given $k$. 
On the other hand, since these subhorizon modes simply redshift, the number density simply dilutes with the expansion for all modes after $a\gg a_*$.
We can therefore infer the later behavior of all relevant momenta modes directly from the relativistic simulations using number conservation instead of explicitly extending them to late times using computationally expensive evaluations of Eq.~(\ref{eqn:generalD}) for the general growth function $D$.  

More specifically, during an epoch where the relevant $q$ modes are still relativistic and $\omega \phi^2/a \approx q \phi^2/a \propto q\phi^2/\tau$, we can define an effective number density
\begin{eqnarray}\label{eqn:neff}
n_{\rm eff}(\bk;\tau) 
&\propto& 
\int \frac{d^3 q}{(2\pi)^3} D(q;\tau) D(|\bk-\bq|;\tau) \frac{\sqrt{ q | \bk-\bq|}}{\tau} \nonumber\\
&&\times 
\phi({\bq};0) \phi({\bk-\bq};0),
\nonumber\\
&\approx& \int \frac{d^3 q}{(2\pi)^3} 
\frac{\sin(q\tau)}{\sqrt{q \tau}} \frac{\sin(|\bk-\bq|\tau)}{\sqrt{|\bk-\bq|\tau}} \frac{1}{\tau^2} \nonumber\\
&&\times 
\phi({\bq};0) \phi({\bk-\bq};0),
\end{eqnarray} 
which then reflects the spectrum of $\phi^2(\bk)$ and their weights in $\phi(\bq)$ {\it after} the waves have become nonrelativistic. 
Notice that given an initial white noise spectrum for $q \lesssim 1/\tau_i$, the field modes are no longer white after free streaming, but the integral in Eq.~(\ref{eqn:neff}) is still dominated by $q \sim 1/\tau_i$ and these modes produce a 
white noise spectrum in $n_{\rm eff}/\langle n_{\rm eff}\rangle$ for $k \lesssim 1/\tau_i$ that is constant in time. Therefore, the white noise power at $k \ll 1/\tau_i$ still comes mainly from momentum modes $q \sim 1/\tau_i$ 
(see discussion of Fig.~\ref{fig:uniform_low_pass}), which as we will see in the next section is analogous to the beat frequency from the superposition of closely spaced high frequency modes.

\begin{figure}
    \centering
    \includegraphics[width=1\linewidth]{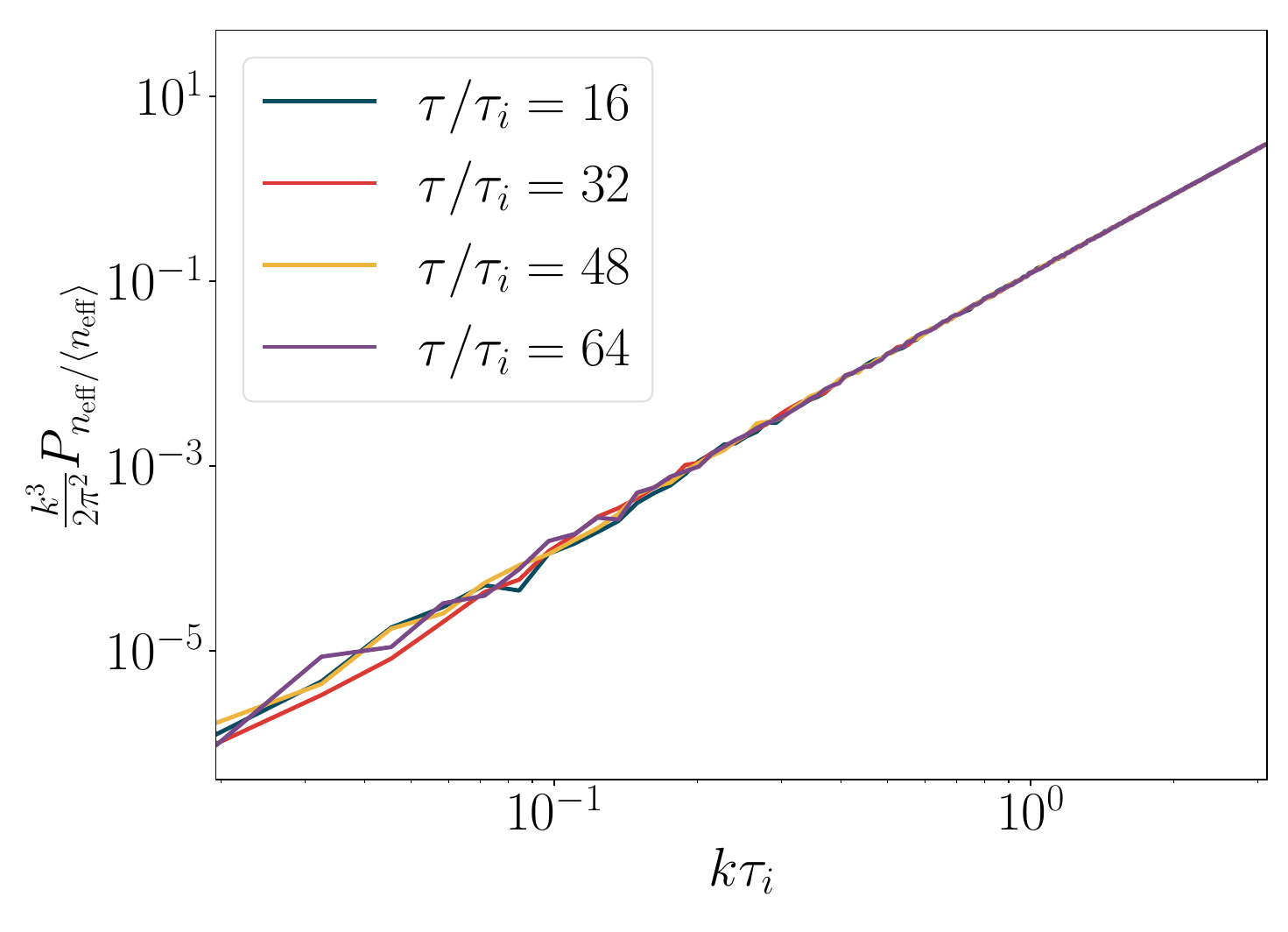}
    \caption{Effective number density power spectrum (see Eq.~\ref{eqn:neff}) for the simulation case of Fig.~\ref{fig:uniform_random_vis}.  Since number density is conserved during the transition of each momentum mode from relativistic to nonrelativistic, the effective number density weights the momentum modes according to the final dark matter density, and its fractional power spectrum remains constant and white at late times despite the free streaming of the underlying modes themselves. }
    \label{fig:spectrum_neff}
\end{figure}

We illustrate this behavior and construction using the random pixel simulations of Fig.~\ref{fig:uniform_random_power} in Fig.~\ref{fig:spectrum_neff}.  Notice that the power spectrum now remains constant and white for the effective number density and now reflects $\phi^2$ after all of the relevant momenta are nonrelativistic.

On the other hand, this constancy of the power spectrum of $n_{\rm eff}/\langle n_{\rm eff}\rangle$ should not be equated with a particle number density in real space since its spectrum is still composed of field modes with different $q$, both in magnitude around $1/\tau_i$ and in direction, due to the incoherent superposition of contribution from individual horizon scale patches, i.e.\ the wave analogue of a phase space distribution.

\section{Incoherent Superposition}
\label{sec:incoherence}

We have seen in the previous section that the effective number density fluctuations in a free scalar field $\phi$, with an initial white noise spectrum from causal production, do not damp by free streaming.  On the other hand, the initial field fluctuations themselves strongly evolve due to free streaming on scales smaller than the free streaming length $\lfs$ in Eq.~(\ref{eqn:lfs}).

As alluded to through visualizations of simulations in the previous section, the resolution of this apparent paradox is that the effective number density fluctuations constructed from $\phi^2(\bx)$ with different field momenta should instead be considered as a phase space number density fluctuation, just as it would be for particle dark matter.

For the case considered in the previous section where the field fluctuations at some $q \gg a m$ are relativistic, the correspondence to relativistic particles or classical waves is direct.  In the classical limit of high photon occupancy, the radiation associated with the photons would be characterized by its electric field ${\bf E}(\bx)$ and the power in radiation by $|{\bf E}|^2(\bx)$, analogously to $\phi^2(\bx)$.    Despite the fact that the electric field contributed by electromagnetic waves of different momenta $q$ always superimpose, observational quantities like the specific intensity do not carry the cross terms of different $q$.   The reason is that the cross terms average away over many cycles of their respective oscillations.   
Moreover, the two-point correlation  of the specific intensity 
$\langle  |{\bf E}|^2(\bx) \rangle \langle
|{\bf E}|^2(\bx')\rangle$
does not carry either the time averaged power of the individual $q$ modes that $\langle  |{\bf E}|^2(\bx)
|{\bf E}|^2(\bx')\rangle$ would: the power spectrum of the time average is not the time average of the instantaneous power.  Phrased in terms of the highly occupied particle states, the rapidly varying fluctuations in the field that are characterized by the $q$-spectrum represent fluctuation in the fine grained phase space or photon occupancy of momentum states.

While the analogy to photons is direct when the $q$-modes of $\phi$ are relativistic, 
this absence of a time averaged effect on $\phi^2$ is also manifest after the $q$-modes of $\phi$ have become non-relativistic but before equality.  Consider the temporal frequency of two non-relativistic modes with $q_1\ne q_2 = |\bk -\bq_1|$
\begin{equation}\label{eqn:omeganr}
\omega_{1,2} \approx ma + \frac{1}{2}\frac{q_{1,2}^2}{ma} .
\end{equation}
The cross term contribution to $\phi^2$ between the two modes evolves as
\begin{equation}
e^{i \int d\tau (\omega_1-\omega_2)} \approx 
e^{i \int d\tau \frac{\bk \cdot \bq_1}{a m}},
\end{equation}
where the approximation assumes $k\ll q_1,q_2$.
As with $\lfs$, this integral grows logarithmically until equality.  In fact, unless the wavelength exceeds the free streaming length $k\lfs(q_1;\tau) \ll 1$, the contribution to $\phi^2$ of the interference between this pair oscillates in time and would prevent the interference from enabling the growth of dark matter density perturbations below the free streaming scale.

\begin{figure}
    \centering
    \includegraphics[width=1\linewidth]{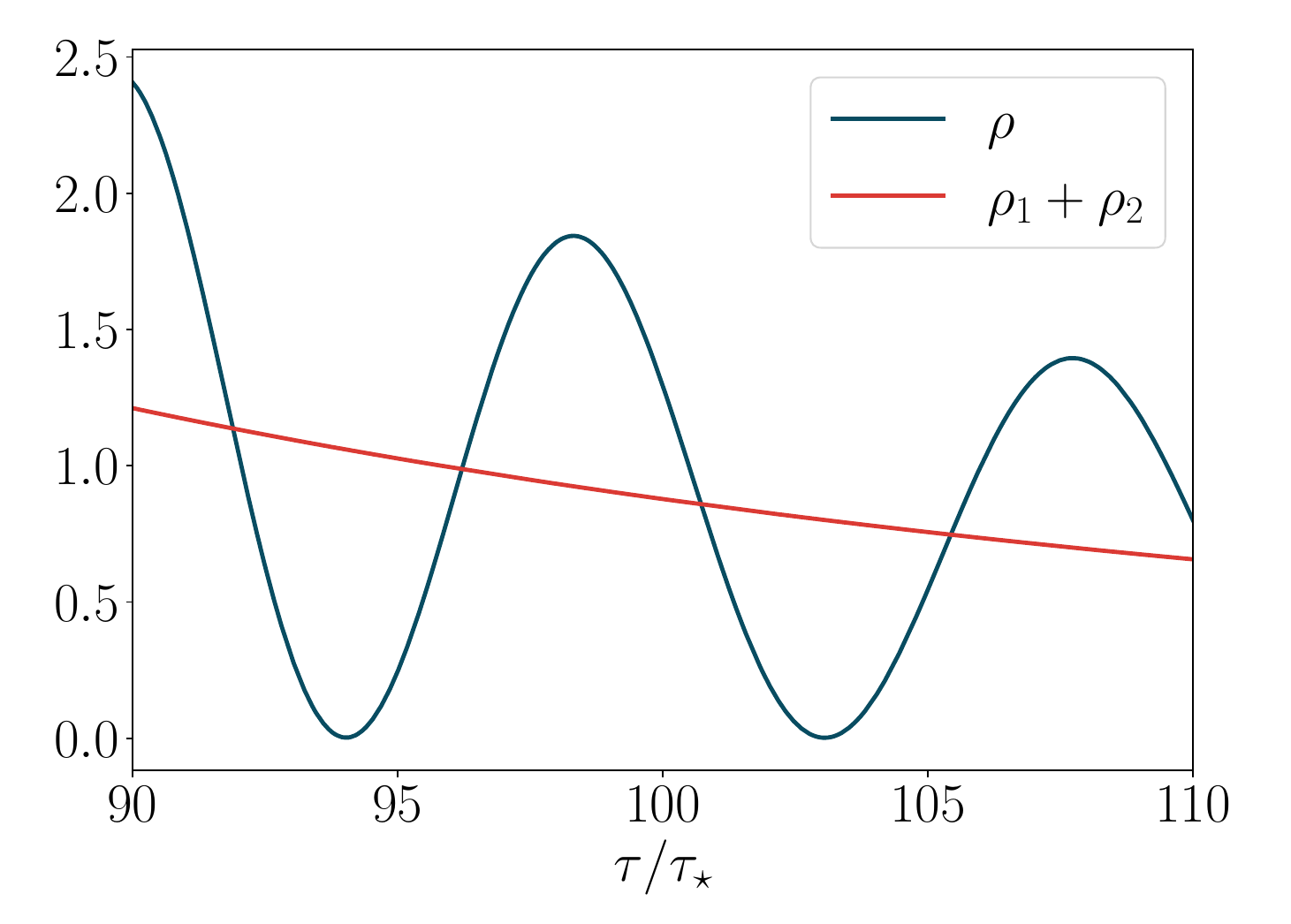}
    \caption{
    Density evolution and interference of two field momentum modes $q_1/a_*m=22,25$ that were relativistic at $\tau_*$ but nonrelativistic at $\tau\sim 100\tau_*$.  The beating of these high frequency modes produces density fluctuations on longer scales and larger wavelengths, but the Hubble time averaged density reflects the incoherent sum of the density contributions of the individual modes $\rho_1+\rho_2$. Normalization is arbitrary.}
    \label{fig:beats}
\end{figure}

To see this explicitly, in Fig.\ \ref{fig:beats} we plot the time evolution of the total density as constructed from just two $q$-modes, $\phi=\phi_1+\phi_2$ with $q_1/a_* m=22$ and $q_2/a_* m=25$ using the full growth function (\ref{eqn:generalD}) for their time evolution from initially equal amplitudes.   The total density, 
\begin{equation}\label{eqn:totaldensity}
\rho = \frac{1}{2} \left( \frac{d\phi}{dt} \right)^2 + 
\frac{1}{2a^2 }(\nabla\phi)^2 + 
\frac{m^2}{2}\phi^2 ,
\end{equation}
contains both quadratic combinations from the same $q$ and the cross combination or impact of the superposition of the two.  Here $\nabla\phi$ is the spatial field gradient in comoving coordinates, though its impact on $\rho$ is subdominant for these non-relativistic modes where $\langle(d\phi/dt)^2\rangle \sim m^2\langle\phi^2\rangle$ on time average.      We compare this to the quadrature sum of the individual modes $\rho_1+\rho_2$ that omits the superposition term.  
Even though each individual field mode oscillates with time according to Eq.~(\ref{eqn:omeganr}), with phase evolution given by $\omega\tau \sim m a\tau  = (\tau/\tau_*)^2 \sim 10^{4}$, $\rho_{1}+\rho_{2}$ only evolves on the Hubble time scale since energy is covariantly conserved between the kinetic and potential terms of each term.
On the other hand, the superposition causes a beat contribution that oscillates faster than the Hubble time scale and much slower than the mass scale, but time averages to 
$\langle \rho \rangle_{\tau} = \rho_1+\rho_2$.    Moreover, fluctuations for a given beat frequency $\bk$ are composed of all possible pairs of high frequency momenta that satisfy $\bk = \bq_1+\bq_2$ and each pair contributes with a random temporal phase.

After equality $a \gg a_{\rm eq}$ and for $a\gg a_*$, we can remove the fast but $q$ independent mass scale oscillations in Eq.~(\ref{eqn:omeganr}) by recasting the scalar field with the Schr\"{o}dinger wavefunction $\psi$
\begin{equation}
\phi = \frac{1}{\sqrt{2}}(\psi e^{-i mt} + \psi^* e^{imt})
\end{equation}
as done with simulations of ultralight dark matter
\cite{Widrow:1993qq}.  In this case, the temporal oscillations between $q_1\ne q_2$ components of $\psi$ are slow but the spatial phases embedded in $\psi$ are still incoherent.

Well above the de-Broglie scale where $k\ll q$, the field $\psi(\bx)$ encodes the full phase space of the collisionless dark matter through the Husimi representation.   This effectively assigns the spatial variation induced by $q$ on scales that are smaller than some spatial smoothing scale $\eta$ to the momentum distribution at $\bp=\bq$:
\begin{eqnarray}\label{eqn:Husimi}
\Psi(\bx,\bp) &=& \left( \frac{ 1}{2\pi}\right)^{3/2}  \left( \frac{1}{\pi \eta^2} \right)^{3/4} 
 \int d^3 r \psi(\br) \\
&&\times\exp\left( -\frac{(\bx -\br)^2}{2\eta^2} -i 
[\bp \cdot(\br -\bx/2) ]\right). \nonumber
\end{eqnarray}
The analog of the phase space distribution function 
\begin{equation}\label{eqn:Husimif}
f(\bx,\bp) \equiv |\Psi(\bx,\bp)|^2
\end{equation}
obeys the collisionless Boltzmann equation.   Notice that spatial fluctuations in the wave $\psi$ encode both the spatial and the momentum variations of the phase space distribution.   Moreover, the spatial variations due to oscillations of a spectrum of $q$-modes produce a multistream phase space distribution where multiple momenta $\bp$ exist at the same position $\bx$.

The relevance of the Husimi representation for our case of effective number density fluctuations from wave interference between different $q$ modes can be more directly seen by reversing the construction.  Starting with a target  particle phase space distribution $f_p$, we can construct the corresponding wave $\psi$ in the same way as the random field simulations on a pixelized grid of the previous section.  In general, given the discrete Fourier transform momenta indexed by $i$, one assigns $\psi$ at grid points indexed by $\iota$ as \cite{Widrow:1993qq}
\begin{equation}
\psi(\bx_\iota) = \sum_i \sqrt{ f_p(\bx_\iota,\bp_i) } 
e^{i \bx_\iota \cdot \bp_i}e^{i\alpha_i} ,
\label{eqn:psifromf}
\end{equation}
where $\alpha_i$ is a random spatial phase for each momentum.  By explicit substitution, Ref.~\cite{Widrow:1993qq} showed  that the Husimi construction (\ref{eqn:Husimi}) returns the phase space distribution $f\approx f_p$ since the cross terms between momenta
$i \ne j$ average away due to spatial phase incoherence. 
Notice that the effective averaging in Eq.~(\ref{eqn:Husimi}) occurs before the squaring in Eq.~(\ref{eqn:Husimif}) and is analogous to the temporal averaging considered in Fig.~\ref{fig:beats} for incoherent temporal phases.
These cross terms are negligible so long as the spatial scale of interest is much longer than the smoothing scale $\eta$ and $f_p$ is smooth or averaged over momenta scales $1/\eta$.

In our case of interest the effective number density fluctuation  is on a scale $k \ll q$ and thus comes solely from the interference of different $q$ modes.  
Here $\psi$ takes the form of
Eq.~(\ref{eqn:psifromf}) where $f_p$ is a function of $\bp$ alone, i.e.\ the incoherent sum of many plane wave fluctuations.   It is then immediately clear that in the Husimi representation
the directional variation of contributions from incoherent sources visualized in the previous section becomes an anisotropy in the phase space distribution $f(\bx, \bp) = f(\bp)$ rather than an inhomogeneity in the spatial distribution $f(\bx,\bp)=f(\bx)$.

   As with particle dark matter, the free streaming of waves produces a random distribution of momentum streams at each spatial point rather than just spatial fluctuations of a cold distribution. 
The white noise effective density fluctuations from interference are indeed preserved by freestreaming but are represented by fine grain phase space fluctuations, and in the coarse grain phase space are represented by the particle velocity dispersion.   This mapping of wave interference fluctuations to particle velocity dispersion has been explicitly studied for the case where an initially cold, single stream distribution becomes multistream \cite{Gough:2022pof} and eventually virializes.\footnote{Our effective density fluctuations are their ``hidden" density fluctuations (their \S 4.2.3 and their Eq.\ (42) for the velocity dispersion or equivalently the quantum pressure of fuzzy dark matter).}
   
Since CDM isocurvature fluctuations grow in the matter dominated epoch, this would cause the relative transfer function to evolve even if the phase space density fluctuations are conserved and not gravitationally unstable.
For \Schro-Poisson simulation based tests of this construction, see Ref.~\cite{Mocz:2018ium}, and for techniques to remove interference density fluctuations from the consideration of gravitationally bound systems, see Ref.~\cite{May:2022gus}.

\section{Discussion}\label{sec:discussion}

We have elucidated the relationship between the free streaming of particle dark matter and wave dark matter and shown how to map the properties of the former onto the latter. 

For axion wave dark matter where Peccei-Quinn symmetry breaking occurs after inflation, axion field fluctuations behave like a warm component of particle dark matter in the sense of possessing a mildly relativistic wave spectrum originated from misalignment and axion string emission.  Correspondingly, the free streaming length and its impact on curvature fluctuations is only larger than that of cold axions from the pre-inflationary scenario by a logarithmic factor.  

We quantify these scaling in terms of the free streaming scale as a function of the characteristic momentum, $\lfs(q_*=a_* m)$, that corresponds to the horizon wavenumber when axions begin their oscillations $H(a_*)=m$, and compare this to the cold axion Jeans scale where the free streaming of wave fluctuations from curvature perturbations overtake their own wavelength.  For axions, free streaming bounds on cold axions or fuzzy dark matter can be roughly translated to the warm case with these scaling relations. 
However, for warm axions from the post-inflationary scenario, isocurvature fluctuations from the random misalignment on the Hubble scale at $a_*$ provide the stronger bound.

If wave dark matter is born ultrarelativistic, then free streaming can have a larger effect as with ``hot" dark matter.  
We provide closed form expressions for the free streaming length $\lfs(q)$ for an arbitrary momentum in Appendix~\ref{sec:lfsscaling} that can be used to assess its impact in any given model with its specific momentum distribution.  Generally in such scenarios, the isocurvature fluctuations from causal generation in horizon scale patches at birth can also be affected by free streaming.  We illustrate the effect on 
phase coherent patches and show that they also rapidly damp via free streaming,  leaving only phase incoherent transient fluctuations from the superposition of waves of different patches in the effective number density.   Despite the free streaming damping of these waves, these incoherent effective number density fluctuations are constant and white at late times when all modes are non-relativistic.

However, these effective number density fluctuations are not spatial number density fluctuations in the wave dark matter, but rather the wave analogue of phase space density fluctuations.   At a given spatial position, these fluctuations are composed of multiple field momentum streams from the incoherent sources and the impact of free streaming is similar to the directional damping of collisionless particles.  While relativistic, the process is the same as the creation of CMB anisotropy out of plasma inhomogeneities before recombination.  As the wave momenta become non-relativistic, the process is analogous to the free streaming damping of fluctuations in massive neutrinos.  

More specifically, we show that these free streamed effective number density fluctuations do not behave like real space number density fluctuations over a Hubble or dynamical time in the background or spatially averaged on scales much larger than the de Broglie wavelength of the momentum components.\footnote{This should be contrasted with structure closer to the de Broglie scale, where wave effects and interactions can lead to the formation of solitons in axion miniclusters \cite{Hogan:1988mp,Kolb:1993zz,Eggemeier:2019jsu}.}  Observables that evolve over a dynamical time respond to the time average of these fluctuations.   During radiation domination, 
these fluctuations oscillate at the beat frequency of the combination of field momenta that compose them; during matter domination, the effective or Husimi phase space representation of the wave dark matter explicitly maps them into multiple momentum streams of the phase space, much like warm or hot dark matter.

Therefore, the astrophysical effects of warm or hot fuzzy dark matter isocurvature fluctuations may also differ from those of cold dark matter isocurvature fluctuations below their free streaming length in a manner that depends on the initial momentum distribution and observable in question.  Beyond the axion case, where the free streaming scale is relatively small, we leave the evaluation of specific models and observables to future simulation work.

\appendix

\section{Free Streaming Scalings}
\label{sec:lfsscaling}

In Eq.~(\ref{eqn:lfsscaled}), we defined the free streaming length $\lfs$   in units of the comoving Hubble length at matter radiation equality $  F(q/a_{\rm eq}m,a/a_{\rm eq}) =a_{\rm eq} H_{\rm eq} \lfs/\sqrt{2}$ of a momentum component $q$ through the free streaming integral in Eq.~(\ref{eqn:lfs}).  In $\Lambda$CDM before dark energy domination, we can explicitly evaluate this integral to obtain
\begin{equation}\label{eqn:lfsexact}
F(\hat q,y) = \frac{\qaem \left[\mathcal{F}(\varphi(\qaem,y), \mu(\qaem)) -\mathcal{F}(\varphi(\qaem,0), \mu(\qaem))\right]}{\left(1 + \qaem^2\right)^{\frac{1}{4}}},
\end{equation}
where $\mathcal{F}(\varphi, m)$ is the incomplete elliptical integral of the first kind with arguments
\begin{eqnarray}
\varphi(\qaem,y)&=& \arccos\left(\frac{\sqrt{1 + \hat{q}^2}-1-y}{\sqrt{1 + \hat{q}^2}+1+y}\right), \nonumber\\
\mu(\qaem) &=&  \frac{1}{2} \left(1 + \sqrt{\frac{1}{1 + \qaem^2}} \right).
\end{eqnarray}

Although the exact free streaming integral is used in all numerical computations in this work, it is useful to examine the approximate scaling behavior of this solution in various regimes of interest and provide a global approximation that is simple to evaluate. 

For $y=a/a_{\rm eq} \ll 1$ the integral is manifestly simple to evaluate and becomes
\begin{equation}
F(\qaem,y) \approx \qaem \sinh^{-1} (y/\qaem), \qquad y\ll1 .
\end{equation}
Notice that for $\qaem \gg y$, the wave is ultrarelativistic $q/am\gg 1$, $F\rightarrow y$, and the free streaming length is the horizon length $\lfs =\tau$ as expected.
This growth continues until $a \sim q/m$ and thereafter $\lfs$ grows logarithmically from its value at $\tau|_{a=q/m}$,  
\begin{equation} 
F(\qaem,y) \approx \qaem \ln\left( 2y/\qaem \right),
\quad \qaem \ll y\ll1.
\end{equation}
This logarithmic growth halts at matter-radiation equality and brings the free streaming scale for modes that are non-relativistic at equality to
\begin{equation} \label{eqn:lfsqy}
F(\qaem,y) \approx 
\qaem \left( \ln\frac{8}{\qaem} - \frac{2}{\sqrt{y}} \right), \quad\qaem\ll  1\ll  y ,
\end{equation} 
which we provided in Eq.~(\ref{eqn:Flimit}) in its leading order ($y\gg 1$) form.   
For axions this limit (\ref{eqn:lfsqy}) applies to essentially the entirety of the number density spectrum as we have explicitly verified by comparing its use to the full expression (\ref{eqn:lfsexact}) to calculate the transfer function as in Fig.~\ref{fig:transfer}.

For evaluating the small contribution from axion momenta that are still relativistic at equality $\qaem >1$ or in more general models,
it is useful to examine late time contributions to the free streaming integral.  Here the free streaming length continues to grow as $\lfs \approx \tau \propto a^{1/2}$ during matter domination until $a= q/m$.   Taking the matter only scaling for $H(a)=H_0 \Omega_m^{1/2} a^{-3/2}$ in the integral (\ref{eqn:lfs}), we find 
\begin{equation}
\lfs = \frac{2 a^{1/2}}{H_0 \Omega_m^{1/2} }
G(q/am)
, \quad \qaem\gg 1
\end{equation} 
where
\begin{eqnarray}\label{eqn:G}
G(f) &=& 
{}_2 F_1[1/4,1/2,5/4,-f^{-2}] \\
&\approx& \frac{K_{1/2} f^{1/2} }
{[1+ 10 f^{1/2}/3 + K_{1/2}^4 f^{2}]^{1/4}},\nonumber
\end{eqnarray}
where ${}_2 F_1$ is the  hypergeometric function, with $K_{1/2}\approx 1.854$
for the complete elliptic integral of the first kind $K_m$
and the approximation is good to a few percent for all $f$.
Notice that the $q\gg am$ limit again returns $\lfs = 2 a^{1/2}/H_0\Omega_m^{1/2} = \tau$ as expected.
On the other hand for modes that have become non-relativistic between $a_{\rm eq}$ and $a$, the free-streaming length approaches a constant value
\begin{eqnarray}\label{eqn:lfsnr}
\lfs &\approx& 2 K_{1/2} \frac{\sqrt{q/m}}{H_0 \Omega_m^{1/2} },
\quad
 a_{\rm eq} \ll q/m \ll a,
\end{eqnarray}
associated with the horizon length at the epoch the momentum becomes non-relativistic.

In summary, to a few percent accuracy we can approximate the
whole solution Eq.~(\ref{eqn:lfsexact}) in the $y\gg 1$ regime by joining these approximations
\begin{align}
F(\qaem,y) =
\begin{cases}
\qaem \ln(\frac{8}{\qaem}), & \qaem < 1 \\
2 y^{1/2} [G(\frac{\qaem}{y}) - G(\frac{1}{y})] + \ln 8, & \qaem \ge 1
\end{cases}
\end{align}
and using the simple approximation for $G$ in Eq.~(\ref{eqn:G}) such that all of the various scaling regimes are manifest.

\begin{figure}
    \centering
    \includegraphics[width=1\linewidth]{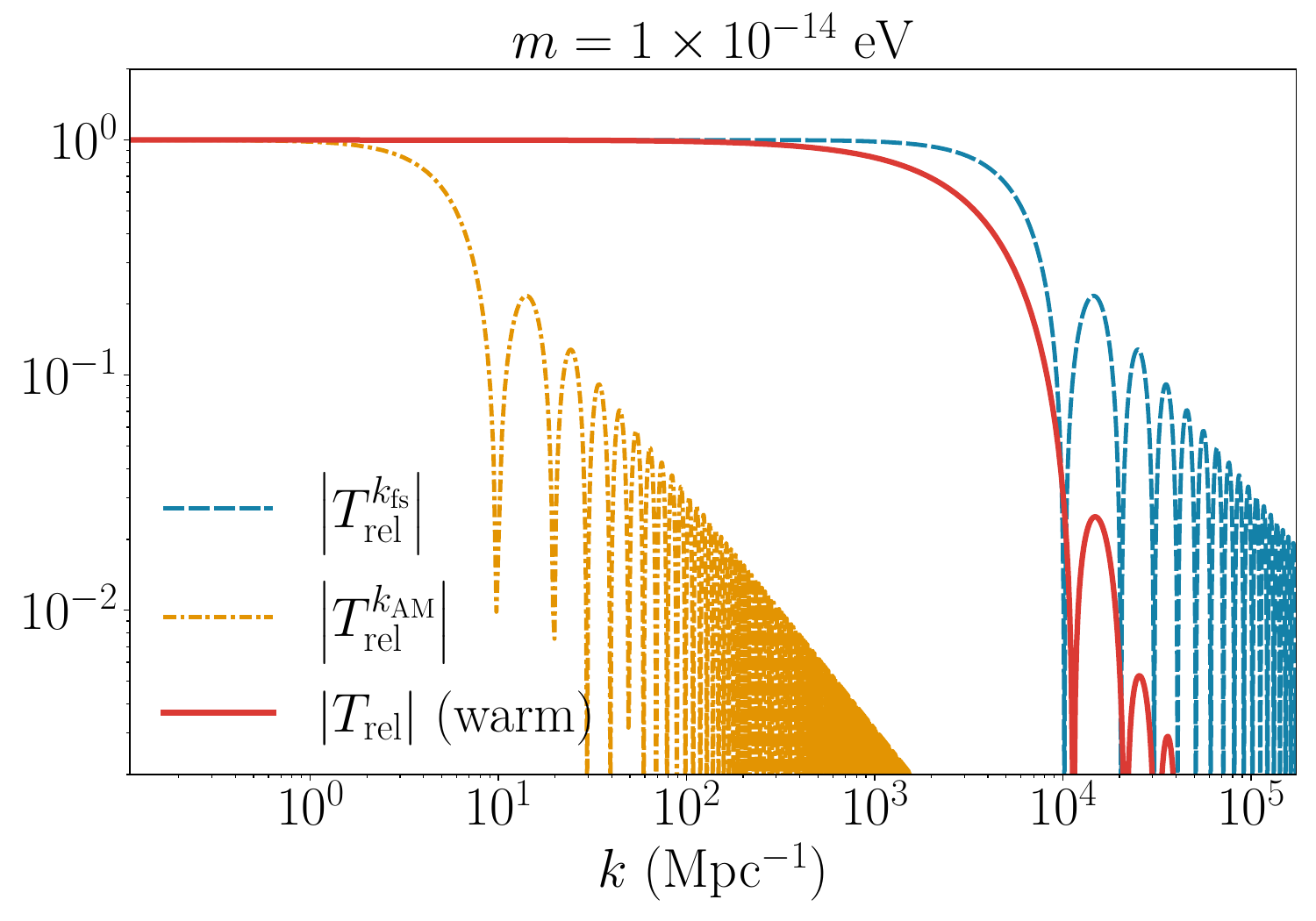}    \caption{Approximations for the relative transfer function (\ref{eqn:transfer}) [$T_{\rm rel}$ warm], using $k_{\rm fs} = \lfs^{-1}(q_*)$ in Eq.~(\ref{eqn:transferapprox})   [$T_{\rm rel}^{k_{\rm fs}}$], and an approximation from
    Ref.~\cite{Amin:2022nlh} v2 using $k_{\rm AM}$
     [$T_{\rm rel}^{k_{\rm AM}}$]
    of Eq.~(\ref{eqn:lfsaverage}) instead. The former captures the scale at which free streaming occurs whereas the latter changes this scale to be orders of magnitude smaller in $k$ and would cause this mass $m=10^{-14}$eV to be inappropriately ruled out. }
    \label{fig:comparison_qfs}
\end{figure}

\section{Averaging over Momenta}
\label{sec:averaging}

In the main text Eq.~(\ref{eqn:transfer}), we averaged the effect of free streaming over the momentum distribution of the axions $q^3 P_\phi(q)$ to approximate the net impact on the transfer function.  For the spectrum of Eq.~(\ref{eqn:Pphi}) which has a sharp peak at $q_*$, the impact is similar to evaluating an effective $k_{\rm fs}= \lfs^{-1}(q_*)$  and taking
\begin{equation}\label{eqn:transferapprox}
T_{\rm rel}^{k_{\rm fs}}(k) = \frac{\sin (k/k_{\rm fs})}{k/k_{\rm fs}}
\end{equation}
in the region $k \sim k_{\rm fs}$ where the damping starts to have its effect.
In Fig.~\ref{fig:comparison_qfs}, we compare this approximation to Eq.~(\ref{eqn:transfer}).  Notice that this approximation does provide the correct scaling for the wavenumbers where free streaming starts to become important but underestimates the effect at higher $k$.  Mathematically, this comes about because Eq.~(\ref{eqn:transfer}) is an integration over an oscillating quantity.  Even in this case where the spectrum is peaked near $q_*$, the phase $k\lfs(q)$ varies over an increasingly large range as $k$ increases.  Note however that the derivation of Eq.~(\ref{eqn:transfer}) itself in Ref.~\cite{Amin:2022nlh} is not ensured to be physically valid for $k\lfs \gg 1$ and should be considered as an estimate for the half power point.

In Ref.~\cite{Amin:2022nlh} v2, this effective $k_{\rm fs}$ approach was adopted but instead of weighting the impact of free streaming  by the number density spectrum, they
equated the Taylor expansion  of 
Eq.~(\ref{eqn:transfer})
\begin{equation}\label{eqn:taylor}
\lim_{k\rightarrow 0}T_{\rm rel}(k) \approx 1 - \frac{1}{{6}}\frac{k^2}{k_{\rm AM}^2}
\end{equation}
where
\begin{equation}\label{eqn:lfsaverage}
    \frac{1}{k_{\rm AM}^2 } = \frac{ \int_0^{a m} d\ln q \frac{q^3}{2\pi^2} P_{\phi}(q) \lfs^2(q)}{\int  d\ln q \frac{q^3}{2\pi^2} P_{\phi}(q)},
\end{equation}
to that of Eq.~(\ref{eqn:transferapprox})
\begin{equation}\label{eqn:taylorapprox}
\lim_{k\rightarrow 0}T_{\rm rel}^{k_{\rm fs}}(k) \approx 1 - \frac{1}{{6}}\frac{k^2}{k_{\rm fs}^2}
\end{equation}
to imply that  $k_{\rm fs} \rightarrow k_{\rm AM}$ in
Eq.~(\ref{eqn:transferapprox}) and $T_{\rm rel}^{k_{\rm fs}} \rightarrow T_{\rm rel}^{k_{\rm AM}}$ (their Eq.\ 5\,v2).
In Fig.~\ref{fig:comparison_qfs}, we compare this transfer function to Eq.~(\ref{eqn:transfer}) and
Eq.~(\ref{eqn:transferapprox}).   Notice that this weighting scheme overestimates the effect of free streaming by three orders of magnitude, with $k_{\rm fs} \approx 3.2\times 10^3$\,Mpc$^{-1}$ while $k_{\rm AM} \approx 3.1 $\,Mpc$^{-1}$. The overestimate is so large that this $m=10^{-14}$eV example would be inappropriately ruled out.
In Ref.~\cite{Amin:2022nlh}\,v2, this approximation was used to place a bound of $m>2\times 10^{-12}$eV for the spectrum considered here and $m >10^{-12}$eV for their axion parameterization with mildly relativistic modes from string decay.

\begin{figure}
    \centering
    \includegraphics[width=1\linewidth]{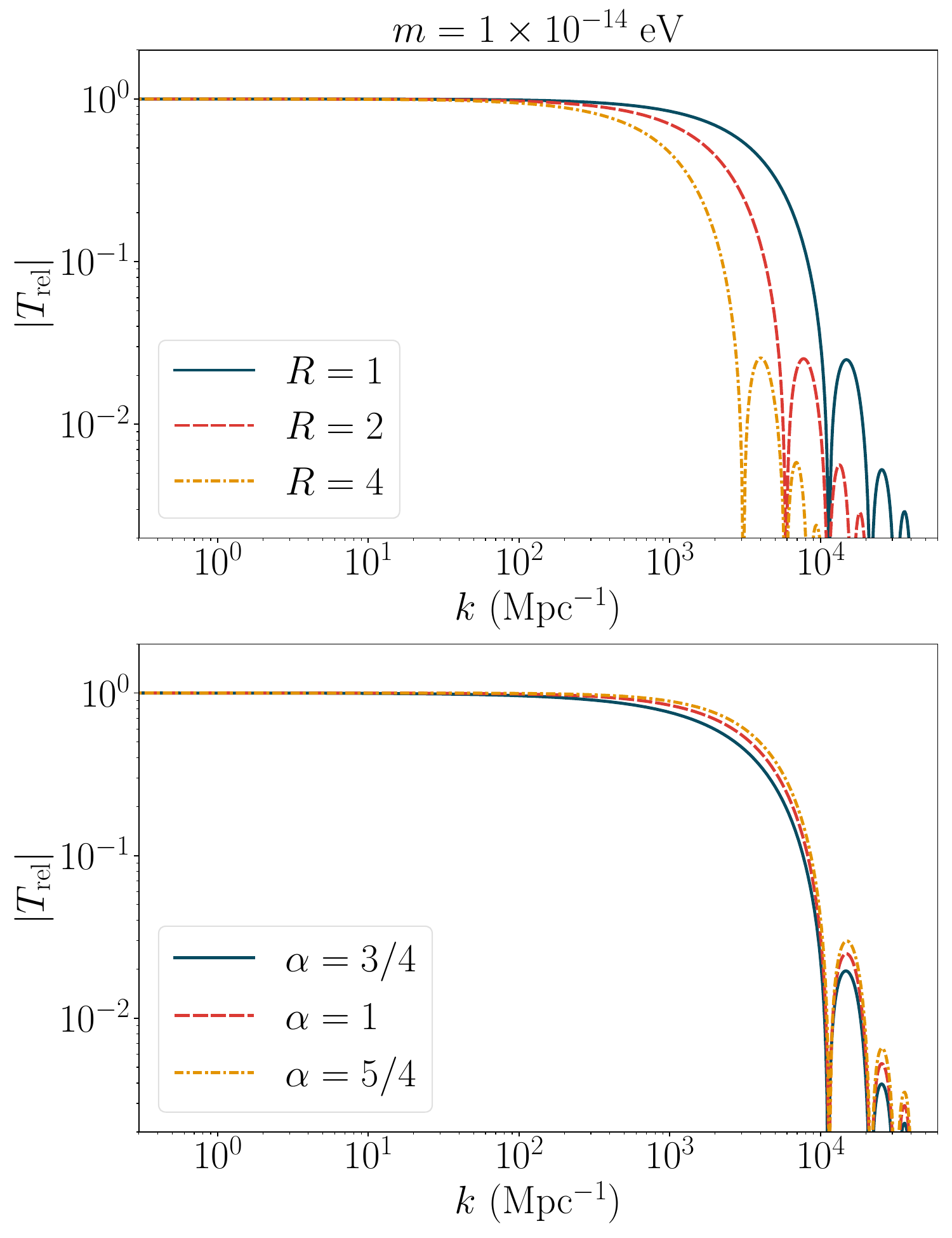}
    \caption{Relative transfer function for the warm axions at the same $m=10^{-14}$ eV mass as Fig~\ref{fig:comparison_qfs} but with variations to the power spectrum which respectively increase the peak momentum by $R=1,2,4$ and alter the high momentum slope $\alpha =3/4,1,5/4$ using Eq.~(\ref{eqn:Pphi_var}). We see that the variation of $R$ shifts the scale of the free streaming damping for the same $\alpha=1$, whereas altering $\alpha$ for the same $R=1$ only makes small changes in the shape of the damping. }
    \label{fig:Trel_R-alpha}
\end{figure}

This overestimate is tied to the behavior of the high momentum tail with the spectrum in Eq.~(\ref{eqn:Pphi}).  In Eq.~(\ref{eqn:lfsnr}) we show that for waves that become nonrelativistic in the matter dominated regime, $\lfs \propto q^{1/2}$.  Thus for the spectrum $q^3 P_\phi \propto q^{-1}$, the integral in Eq.~(\ref{eqn:lfsaverage}) receives nearly constant contributions per $d\ln q$ up to the $q\approx am$ limit where the waves are still relativistic at the evaluation epoch despite the highly suppressed number of axions with such momenta.
The result is that the estimate of the effective free streaming wavenumber $k_{\rm AM}$ produces a suppression of the transfer function to much smaller wavenumbers or much larger scales than calculated from Eq.~(\ref{eqn:transfer}) or estimated by Eq.~(\ref{eqn:transferapprox}).   Mathematically, the Taylor expansion (\ref{eqn:taylor}) is not a good approximation at $k \sim k_{\rm AM}$ since the dominant momenta near $q_*$ have vanishingly small free streaming effect as $k\rightarrow 0$, causing the second derivative of $T_{\rm rel}$ to run strongly with scale.

The source of this discrepancy is the difference in the weighting scheme.
 Since $\lfs$ grows as $\tau$ for relativistic momenta, the weighting in Eq.~(\ref{eqn:lfsaverage}) allows a very small number density in high momentum waves to dominate the effective free streaming length $k_{\rm AM}$ of the whole population, whereas physically free streaming implies that instead this small component is smooth across scales where the dominant component remains clustered, similar to the small admixture of massive neutrinos and cold dark matter in $\Lambda$CDM.  That both momenta can be represented by the single field $\phi(\bx)$ is also related to the Husimi phase space construction discussed in \S \ref{sec:incoherence}.  The spatially smooth and clustered components are embedded in the inferred momentum distribution.  

Since the key quantity that controls the free streaming effect is the shape of the momentum distribution, 
we also explore variations from Eq.~(\ref{eqn:Pphi}) that adjust the position of the peak in the spectrum and the power law decline from the peak, parameterized by $R$ and $\alpha$ as follows:
\begin{equation}\label{eqn:Pphi_var}
q^3 P_\phi(q) \propto \left( \frac{q}{R q_{*}}\right)^3 \theta(R q_{*}-q) + \left( \frac{R q_*}{q} \right)^\alpha \theta(q-R q_*) .
\end{equation}
Fig.~(\ref{fig:Trel_R-alpha}) shows the corresponding change in the transfer function.
In the top panel, we fix $\alpha = 1$ and increase $R$ from $1$ to $2$ and $4$. The damping wavelength increases nearly linearly with $R$ in accordance with the expectation that the free streaming length scales as $\lfs(Rq_*)$ discussed in \S \ref{sec:sims}.  Varying $\alpha$ in the range where most of the particles still have momenta $\sim q_*$ has a much smaller effect since only the small tail of high momenta waves are affected.  These variations in $\alpha$ encompass the full range found in current state of the art axion simulations \cite{Buschmann:2019icd,Buschmann:2021sdq,Gorghetto:2020qws}.

\acknowledgments

We thank Mustafa Amin, Christian  Capanelli, Keisuke Harigaya, Austin Joyce, Andrew Long, Marilena LoVerde, Evan McDonough for useful conversations. R.L. \& W.H. are supported by U.S. Dept.\ of Energy contract DE-FG02-13ER41958 and the Simons Foundation. HX is supported by Fermi Research Alliance, LLC under Contract DE-AC02-07CH11359 with the U.S. Department of Energy.

\vfill

\bibliographystyle{apsrev4-2}
\bibliography{wfdm.bbl}

\end{document}